\documentclass[pra,twocolumn,showpacs,letterpaper,showpacs,superscriptaddress]{revtex4}
\usepackage{graphicx,amsmath,amssymb,amsfonts,latexsym,color,dcolumn,bm,epsfig,subfigure}
\usepackage[plainpages=false,hyperfootnotes=false,colorlinks=false]{hyperref}
\usepackage[normalem]{ulem}
\usepackage{dsfont}

\newcommand{\abs}[1]{\left\vert#1\right\vert}

\newcommand{\arccot}[0]{\text{arccot}}
\newcommand{\arcsinh}[0]{\text{arcsinh}}

\newcommand{\arccoth}[0]{\text{arccoth}}
\newcommand{\op}{\omega_{\rm p}}
\newcommand{\kb}{k_{\text{B}}}
\newcommand{\vf}{v_{\text{F}}}
\renewcommand{\imath}{\mathrm{i}}

\begin{document}

\title{Quantum thermodynamics of overdamped modes in\\
local and spatially dispersive materials}

\author{D. Reiche}
\affiliation{Humboldt-Universit\"at zu Berlin, Institut f\"ur Physik, AG 
			Theoretische Optik \& Photonik, 12489 Berlin, Germany}
\affiliation{Max-Born-Institut, 12489 Berlin, Germany}
\thanks{Corresponding author: reiche@physik.hu-berlin.de}

\author{K. Busch}

\affiliation{Humboldt-Universit\"at zu Berlin, Institut f\"ur Physik, AG 
			Theoretische Optik \& Photonik, 12489 Berlin, Germany}
\affiliation{Max-Born-Institut, 12489 Berlin, Germany}
\author{F. Intravaia}

\affiliation{Humboldt-Universit\"at zu Berlin, Institut f\"ur Physik, AG 
			Theoretische Optik \& Photonik, 12489 Berlin, Germany}


\begin{abstract} 
The quantum thermodynamical properties of (quasi-normal) overdamped electromagnetic modes (eddy currents) are investigated in the context of the magnetic Casimir-Polder interaction. The role of the material response in terms of spatially local and nonlocal material models is discussed. 
In particular, the focus is set on the system's entropy in the limit of low temperatures. In specific circumstances the spatially local (Drude) model reveals an ``entropy defect'', while spatial dispersion leads to a more regular behavior. We present a detailed description of this phenomenon and of the different mechanisms at work in the system with regard to the eddy modes' properties. Extensively discussing classical and quantum features, we relate our results to the wide range of literature and draw intriguing connections to seemingly distant fields as, e.g., the theory of magnetohydrodynamics and superconductivity.
\end{abstract}

\maketitle


\section{Introduction} 

The interaction of a single microscopic object with the surrounding electromagnetic field is one of the oldest and still one of the most frequently considered problems in physics. In recent years there has been remarkable experimental progress concerning the trapping and manipulation of atoms and small objects, thus opening the way to a variety of novel applications for both engineering and foundational research. 
These applications range from the investigation of dark matter  \cite{Rider16}, the technology of atom-chips and their interest for quantum sensing and computation \cite{Keil16}, all the way down to the novel frontier of so-called \textit{atomtronics} \cite{Amico17}, where Bose-Einstein condensates are shaped to circuits consisting of coherent matter waves \cite{Ryu15,Bell16}.
In all these cases the attention of theoretical investigations is drawn to the interaction between atoms and the (quantized) electromagnetic field, especially in the presence of nano- or micro-structured photonic environments.

Among the most relevant phenomena characterizing such systems we find the Purcell \cite{Purcell46} and the Casimir-Polder effect \cite{Casimir48a,Dzyaloshinskii61}. Qualitatively, these phenomena can be understood as a (complex) shift in the  atomic transition frequencies due to a boundary-condition-induced change in the electromagnetic local density of states. While the Purcell effect originates from the imaginary part of this shift, the Casimir-Polder effect is associated with the corresponding real part. In thermal equilibrium the two effects are related to one another: Passivity implies that real and imaginary part of the frequency shift are connected through the Kramers-Kronig relations. Recently, the Purcell effect, which is a narrow-band phenomenon centered around the atomic transition energy, has been  analyzed in terms of a spectral decomposition with the aim to establish a connection with the system's natural resonances (also addressed as quasi-normal modes -- see for example Refs. \cite{Lalanne18,kristensen19a} for recent reviews). The Casimir-Polder interaction, instead, is a broad-band effect. Therefore, it is usually less sensitive to a modification of the system's resonances and rather depends on the full spectrum of excitable modes.

In this paper we show, however, that an analysis in terms of resonances is also very useful for investigating the Casimir-Polder effect and can help to understand certain quantum thermodynamical characteristics occurring in dissipative systems. 
Indeed, theoretical investigations have shown that in specific cases the entropy linked to the Casimir interaction between two metallic surfaces \cite{Casimir48,Dzyaloshinskii61} is nonzero and depends on the system's parameters in the limit $T\to 0$, contrary to what is usually expected from the third law of thermodynamics, also known as the Nernst theorem \cite{Bezerra04,Intravaia08,Klimchitskaya09b,Klimchitskaya09c,Milton11b}. For a spatially local description of the metal, this behavior was connected with the (quantum) thermodynamic properties of \emph{overdamped} electromagnetic resonances \cite{Intravaia09,Intravaia10a,Guerout14,Guerout16}, which in classical physics are known under the name of eddy (or Foucault) currents \cite{Maxwell71,Jackson75}. 
The same anomaly was reported for the \emph{magnetic} Casimir-Polder interaction between an atom and a metallic surface \cite{Haakh09}. 
Other results showed, however, that a more accurate description of the metal, which includes spatial dispersion and additional dissipation channels such as Landau damping \cite{Landau46}, ``regularizes'' the behavior of the entropy in the two plate configuration \cite{Svetovoy05}.

In the present manuscript we complete the picture sketched above and show that spatial dispersion has the same effect on the Casimir-Polder interaction. Moreover, we investigate the interplay between Landau damping and eddy currents, which enables us to provide a detailed understanding of the energy, frequency and length scales at work in the system. We study the impact of the statistical assumptions used to model the electronic distribution in the (metallic) body and draw a connection to the system's ground state. 

The paper is structured as follows. In Sec. \ref{sec:Model} we describe the physical set-up and give a brief introduction to features of the magnetic Casimir-Polder interaction. Section \ref{nonlocality} is devoted to the material model we use in order to include spatial dispersion in the system. 
We further illustrate how this phenomenon impacts the Casimir-Polder free energy as a function of the atom-surface separation.
In Sec. \ref{EntropyPhenomenology} we review the basics of the Casimir entropy, highlighting the properties which are independent of the configuration under examination, and, as a concrete example, we consider in detail the magnetic Casimir-Polder interaction. 
We report from a phenomenological point of view on the different outcomes of the zero-temperature entropy calculations using, respectively, spatially local and nonlocal material models. 
Further, we discuss the broader connection of our calculations to related theoretical and experimental literature and the Nernst theorem.
The main results of this paper are described in Secs. \ref{LocalEddyModes}, \ref{Defect}, \ref{NonlocalEddyModes} and \ref{SingleMode}. 
Based on an analysis of the system's mode structure in the complex frequency plane, we consider the thermodynamic properties of dispersive (quasi-normal) eddy modes and study their role in the low-temperature characteristics of the atom-surface interaction. We highlight the role that eddy currents play in the determination of the system's entropy and provide a physical picture by investigating the thermodynamics of a single overdamped mode.
In Sec. \ref{MHD} we establish a connection between our results and magnetohydrodynamics as well as London's theory of superconductivity.
We conclude our presentation with a summary and discussion in Sec. \ref{Conclusions}.


\section{Magnetic Casimir-Polder interaction}
\label{sec:Model}
\begin{figure}[!t]
  \centering
    \includegraphics[width=0.49\textwidth]{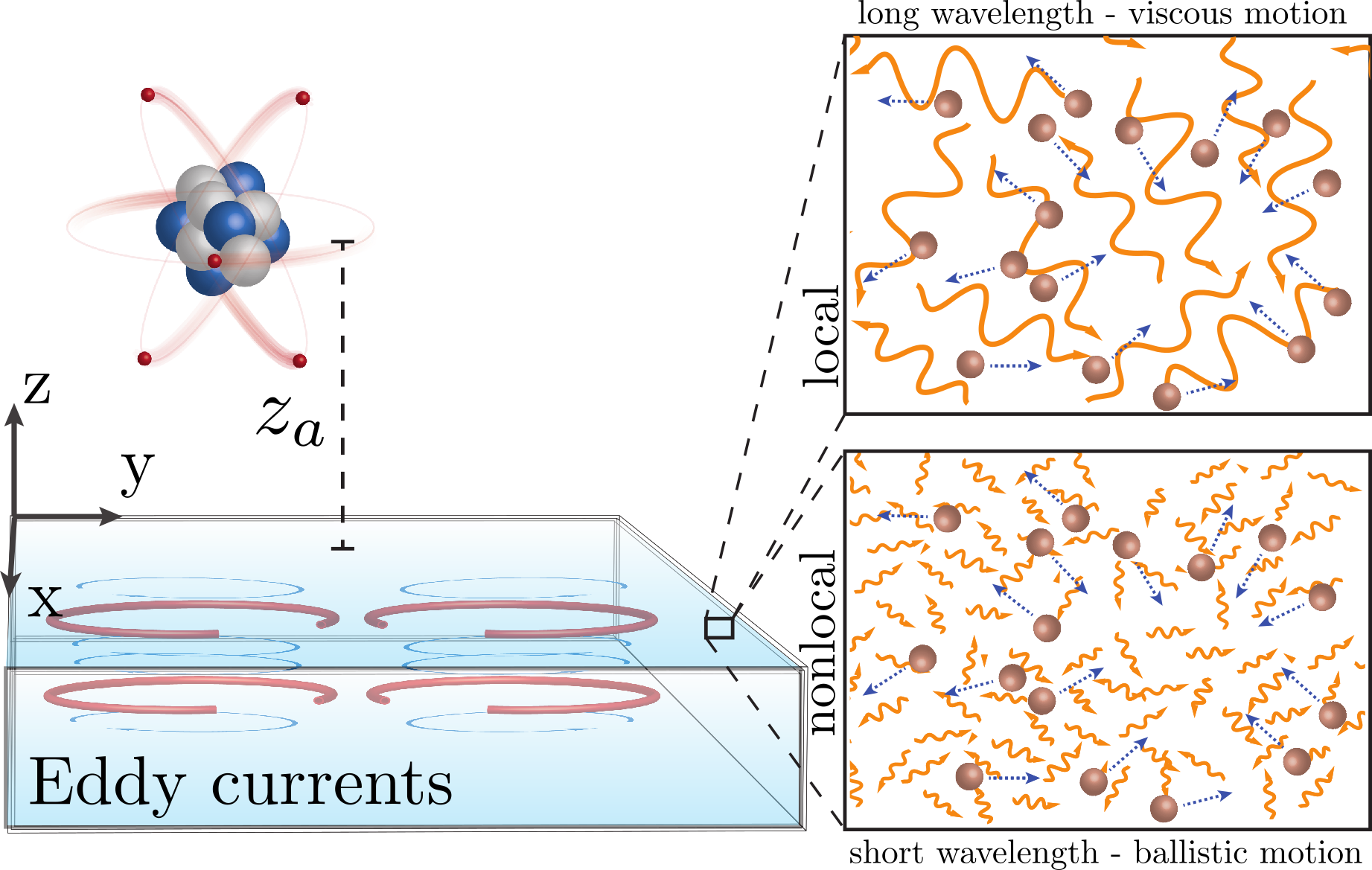}
    \caption{(Color online) Schematic visualization of the set-up we study: A 
    conducting bulk material fills the lower half space 
    (surface coincides with $x$-$y$-plane) and features 
    intrinsic noise due to quantum and thermal fluctuations. 
    In the upper half plane an atom is located at fixed altitude $z_a$ above the 
    surface. Both subsystems, the atom and the bulk material, interact via the 
    quantized electromagnetic field. 
    The electromagnetic response of the metal is described using 
    either a spatially local or a nonlocal material:
    Nonlocality is pronounced at wavelengths or separations shorter than 
    the bulk electron's mean free path.}\label{fig:set-up}
\end{figure}

Perhaps better known for its non-retarded limit, the van der Waals force, the Casimir-Polder effect is a quantum-mechanical force acting on the atom along the direction perpendicular to the surface. It can be understood as the interaction between the fluctuating atomic dipole and its image below the surface, in the bulk material. Therefore it is not surprising that, in addition to the commonly studied electric dipole interaction, there always exists a magnetic contribution to the force.
The magnetic interaction generates a repulsive and much smaller force (typically weaker by a factor $\alpha_{\rm fs}^2$, where $\alpha_{\rm fs}\sim 1/137$ is the fine-structure constant) than the attractive electric dipole interaction \cite{Haakh09b}. 
This magnetic contribution is usually neglected in the investigation of the atom-surface interaction. Nevertheless, the magnetic part reveals intriguing thermodynamical features especially in the low-temperature regime and will be the starting point of our considerations.

Theoretically, the \emph{magnetic} Casimir-Polder free energy can be described in analogy to its electric counterpart \cite{Intravaia11}. For an atom in vacuum at distance $z_a>0$ from a surface placed at $z=0$ (see Fig. \ref{fig:set-up}), the free energy reads
\begin{align}
\label{FreeEnergy}
	\mathcal{F}=-\hbar\int_{0}^{\infty}\frac{{\rm d}\omega}{2\pi}\coth
		\left[\frac{\hbar\omega}{2k_{\rm B}T}
		\right]
		\mathrm{Im}\text{Tr}	
		\left[\underline{\beta}^T(\omega)\cdot\underline{h}^{T}(z_a,\omega)
		\right],
\end{align}
where we assume the entire system to be in equilibrium at temperature $T$. In addition, $\hbar$ is the reduced Planck constant, $k_{\rm B}$ the Boltzmann constant, $\text{Im}[\cdot]$ gives the imaginary part of an expression and $\text{Tr}[\cdot]$ evaluates the trace with respect to spatial coordinates.
The quantity $\underline{\beta}^T(\omega)$ represents the atom's magnetic polarizability tensor in thermal equilibrium and the geometry and material properties of the surface are encoded in the scattered part of the magnetic Green tensor $\underline{h}^{T}(z_a,\omega)$. The superscript in these two last quantities indicates that, in general, they can vary with the temperature $T$. 
In the limit $T\to 0$, the magnetic polarizability $\underline{\beta}^T(\omega)$ reduces to the ground-state polarizability $\underline{\beta}^0(\omega)$. Although the atom possesses a discrete set of (many) eigenstates, in our examples we limit the description to a two-level system for which
\begin{subequations}
\begin{gather}
	\underline{\beta}^T(\omega)=\tanh\left[\frac{T_{a}}{T}\right]
	\underline{\beta}^0(\omega), \\ 
	\beta^0_{ij}(\omega)=\frac{2}{\hbar}\mu_{i}\mu_{j}
	\frac{\omega_{a}}{\omega_{a}^2-(\omega+\imath0^+)^2}.
\end{gather}
\end{subequations}
Here, $\omega_{a}$ characterizes the transition from the ground state to the excited state (usually of the order of a few hundred MHz or a few GHz) and $T_a=\hbar\omega_{a}/(2\kb)$ is the corresponding characteristic atomic temperature ($T_{a}\sim 10^{-2}$K). The constants $\mu_{i,j}$ are the matrix elements between the ground and the excited state of the magnetic dipole vector operator.
The analytical form of the Green tensor for the given configuration has been found by several authors before \cite{Wylie85,Tomas95,Henkel99}.
Upon employing the system's rotational invariance with respect to the $z$-axis, we have at the position of the atom that \cite{Haakh09b,Haakh12}
\begin{align}\label{RIGreen}
	\underline{h}^{T}(z_a,\omega)
		&=\frac{\mu_0}{8\pi}
			\int_{0}^{\infty}{\rm d}p~p\kappa e^{-2\kappa z_a}
			\left[2\frac{p^2}{\kappa^2}r^{\rm TE}(\omega,p)\mathbf{z}\mathbf{z}
			\right.\\\nonumber
		&+\left.
			\left(r^{\rm TE}(\omega,p)
				+\frac{\omega^2}{c^2\kappa^2}r^{\rm TM}(\omega,p)
			\right)(\mathbf{x}\mathbf{x}+\mathbf{y}\mathbf{y})
		  \right],
\end{align}
where $\mu_0$ is the vacuum permeability, $\kappa=\sqrt{p^2-\omega^2/c^2}$ ($\mathrm{Re}[\kappa]>0$ and $\mathrm{Im}[\kappa]<0$) and $p=|\mathbf{p}|=\sqrt{p_x^2+p_y^2}$ is the parallel component of the incident wave vector $\mathbf{k}=p_x\mathbf{x}+p_y\mathbf{y}+q\mathbf{z}$ with $\mathbf{x},\mathbf{y},\mathbf{z}$ being the unit vectors in the direction of the coordinate system. 
For the considered planar geometry, the response of the material is encoded in the transverse electric and magnetic reflection coefficients, $r^{\rm TE}$ and $r^{\rm TM}$, respectively. 
These can be expressed in terms of the transverse electric and magnetic surface impedances, $Z^{\rm TE}$ and $Z^{\rm TM}$ \cite{Jackson75,Ford84}, and read
 \begin{align}\label{reflectioncoefficients}
 	 r^{\rm TE}
 	 	&=\frac{Z^{\rm TE}/Z^{\rm TE}_0-1}{Z^{\rm TE}/Z^{\rm TE}_0+1} 
 	 	&r^{\rm TM}=\frac{1-Z^{\rm TM}/Z^{\rm TM}_0}{1+Z^{\rm TM}/Z^{\rm TM}_0},
 \end{align}
where $Z^{\rm TE/TM}_0$ denote the corresponding vacuum impedances. 
Theoretically, a careful derivation of the surface impedances from first principles is problematic. 
Different models have been developed with the purpose of reconnecting the interface properties to the dynamics of the charge carriers inside the material.
One of the most successful approaches, the semi-classical infinite barrier model (SCIB), takes into account additional boundary conditions on the symmetry properties of the electromagnetic field and the behavior of the electrons at the vacuum-material interface, i.e. specular reflection \cite{Kliewer67,Feibelman82,Ford84,Intravaia15a}.
Assuming specular reflection of electrons at the metal/vacuum interface is one of the simplest boundary conditions one can impose on the electrons' dynamics in order to solve the electromagnetic problem of the spatially nonlocal conducting half space. 
For polycrystalline metals and/or if the root-mean-square roughness of the surface becomes comparable to the bulk electron's mean free path, the assumption of specular reflection might break down and diffusive scattering processes or other approaches should be considered \cite{Kliewer70,Kaganova01,Esquivel04}.
Upon employing the SCIB model, the surface impedances can be written as 
\begin{subequations}
\label{surfaceimpedance}
\begin{gather}
	\label{surfaceimpedanceP}
	Z^{\rm TM}(\omega,p)=\frac{2\imath c}{\pi\omega}
		\int_{0}^{\infty}\frac{{\rm d}q}{k^2}
		\left[\frac{q^2}{\epsilon_t(\omega,k)-\frac{c^{2}k^2}{\omega^{2}}}
			+\frac{p^2}{\epsilon_l(\omega,k)}
		\right],
\\
\label{surfaceimpedanceS}
	Z^{\rm TE}(\omega,p)=\frac{2\imath c}{\pi\omega}
		\int_{0}^{\infty}{\rm d}q~
		\frac{1}{\epsilon_t(\omega,k)-\frac{c^{2}k^2}{\omega^{2}}},
\end{gather}
\end{subequations}
where $c$ is the speed of light and $k=|\mathbf{k}|$.
These surface impedances have been used to describe phenomena occurring at the interface with a metal when the electron's mean-free path cannot be neglected with respect to the spatial variation of the electromagnetic field (e.g. the anomalous skin effect \cite{Pippard47,Reuter48,Chambers52,Kliewer68}). The functions $\epsilon_{l}(\omega,k)$ and $\epsilon_{t}(\omega,k)$ appearing in the previous expressions are the longitudinal and transverse bulk permittivities of the material and quite generally they depend on the frequency and the wave vector of the incoming radiation. The appearance of these quantities in the expressions for the surface's optical response can be understood within the broader context of the linear response and transport theories, where susceptibilities (and by extension the permittivity) are
often recurring central quantities \cite{Ziman92}.
In the local approximation the $k$ dependence is disregarded and if $\epsilon_l(\omega,k)=\epsilon_t(\omega,k) \equiv \epsilon(\omega)$, 
we have
\begin{subequations}
\label{SurfImpadance}
\begin{align}
	\frac{Z^{\rm TM}(\omega,p)}{Z_{0}^{\rm TM}(\omega,p)}
		&=\frac{\sqrt{\frac{\omega^{2}}{c^{2}}\epsilon(\omega)-p^2}}
			{\epsilon(\omega)\sqrt{\frac{\omega^{2}}{c^{2}}-p^2}},\\
	\frac{Z^{\rm TE}(\omega,p)}{Z_{0}^{\rm TE}(\omega,p)}
		&=\frac{\sqrt{\frac{\omega^{2}}{c^{2}}-p^2}}{\sqrt{\epsilon(\omega)
			\frac{\omega^{2}}{c^{2}}-p^2}} 
\end{align}
\end{subequations}
from which we recover the usual Fresnel expressions for the reflection coefficients \cite{Jackson75}.
Finally, it is important to mention the limitations of the previous approach. Treating the material-vacuum interface as a sharp boundary, both in the local and nonlocal case, assumes that the electrons' dynamics stays the same from inside the bulk up to a position infinitesimally close to the surface. This approximation as well as the continuous medium description adds some constraints to the validity of our description, in particular for distances very close to the surface \cite{Klimchitskaya07,Sernelius07}. These might impact the reflection coefficients and therefore affect the prediction on the Casimir(-Polder) force (see also Sec. \ref{Sec:PDControversy}).
Nevertheless, as we highlight below, not only the surface's but also the bulk's properties can play a significant role in determining the behavior of the Casimir-Polder interaction.


\section{Spatial Locality and nonlocality in the material description}
\label{nonlocality}

The previous treatment in terms of surface impedances allows to go beyond the local approximation and for the inclusion of spatial dispersion in the description of our system. It is indeed sufficient to provide expressions for the functions $\epsilon_{l,t}(\omega,k)$ in order to fully characterize the electromagnetic scattering at the vacuum-material interface. Such expressions are derived from the microscopic (quantum) properties of the material. Depending on the chosen approach, $\epsilon_{l,t}(\omega,k)$ can either be quite involved or relatively simple.
Typically, simplicity comes at the cost of an incomplete description of the physical processes occurring in the system. Then, however, the question arises whether these processes are important for the effect under study or not.
The approach we follow tries to accomplish a trade-off between these two aspects.

In metals, nonlocality has been analyzed from several viewpoints and, depending on the accuracy of the description, several models have been developed (see for example \cite{Feibelman82}).
From now on, we describe the material in terms of a plasma of electrons with a statistical distribution $f$ in the phase space, whose dynamics is prescribed by the Boltzmann equation 
\begin{equation}
	\partial_{t}f+\mathbf{v}\cdot\bm{\nabla}_{\mathbf{r}}f
		+\partial_{t}\mathbf{\Pi}\cdot\bm{\nabla}_{\mathbf{\Pi}}f
		=-\Gamma(f-f_{0}).
\label{BoltzmannEq}
\end{equation}
We implicitly assume that, to a good approximation, the momentum $\bm{\Pi}$ of the electrons can be written in terms of their velocity $\mathbf{v}$, i.e.  $\mathbf{\Pi}=m_{e} \mathbf{v}$, where $m_{e}$ denotes the electron mass. This approach shows several similarities with the description of a non-relativistic plasma \cite{Manfredi05}. 
On the r.h.s. of Eq. \eqref{BoltzmannEq} we have written the collision term in the relaxation time approximation by introducing the phenomenological parameter $\Gamma$ \cite{Bhatnagar54}.
We consider small deviations of the actual distribution function $f(\mathbf{r},\mathbf{\Pi},t)$ from the equilibrium distribution $f_{0}(\mathbf{r},\mathbf{\Pi})\equiv f_{0}(v)$, which we assume to depend on the modulus of the velocity $v=\abs{\mathbf{v}}$ only. In equilibrium the motion of electrons is isotropic.
Additionally, we consider that we are dealing with a degenerate plasma of fermions at temperature $T$ much smaller than the Fermi temperature $T\ll T_{\rm F}$ (for a metal $T_{\rm F}\sim 10^{5}$ K). Notice that this is consistent with most of the discussions presented in the present paper, since we are mostly concerned with the limit $T\rightarrow 0$.
Hence, we can express $f_{0}(v)$ in terms of the zero-temperature Fermi-Dirac distribution, $f_{0}(v)=n_{0}/(\frac{4\pi}{3}v_{\rm F}^{3})\theta(v_{\rm F}-v)$, where $v_{\rm F}$ is the Fermi velocity of the electrons in the metal, $n_{0}$ is the equilibrium density of electrons \cite{Manfredi05} and $\theta(x)$ denotes the Heaviside step function. Consequently, we recover the expressions used in the so-called Boltzmann-Mermin model \cite{Kliewer67,Ford84}
\begin{subequations}
\label{permittivities}
 \begin{align}
	\epsilon_l(\omega,k)
		&=1+\frac{\omega_{\rm p}^2}{\omega+\imath\Gamma}
			\frac{3u^2 g_l(u)}{\omega+\imath\Gamma g_l(u)},\\
	\epsilon_t(\omega,k)
		&=1-\frac{\omega_{\rm p}^2}{\omega(\omega+\imath\Gamma)}g_t(u),
 \end{align}
 \end{subequations}
where $\omega_{\rm p}$ is the plasma frequency of the metal and
\begin{subequations}
\begin{gather}
 g_l(u)=1-u \, \arccoth\left(u\right)\\
 g_t(u)=\frac{3}{2}\left[u^{2}-(u^2-1)u \, \arccoth\left(u\right)\right]
\end{gather} 
  \end{subequations}
are dimensionless functions of $u=(\omega+\imath\Gamma)/(v_{\rm F} k)$. 
It is important to underline that, despite the equation of motion in Eq. \eqref{BoltzmannEq} has a classical origin, we are including some ``quantumness'' in the system through the Pauli exclusion principle by considering the zero temperature Fermi-Dirac distribution as equilibrium distribution. The dynamics of the plasma is effectively modified through the addition of a pressure term behaving at low frequency as the Thomas-Fermi pressure. Further,
Eqs. \eqref{permittivities} coincide with the semiclassical limit of the Lindhard-Mermin dielectric functions \cite{Pines66,Lindhard54,Mermin70,Ford84,Dressel02,Note1}, which incorporates quantum corrections as soon as the relevant wave vectors become comparable to or larger than the Fermi wave vector $k_{\rm F}=m_{e} v_{\rm F}/\hbar$. In other words, as long as the wavelength of the radiation is much larger than the de Broglie wavelength of the electron at the Fermi surface our semiclassical descriptions holds. In terms of the parameters of our system this is equivalent to saying that the quantum correction would become relevant for atom-surface separations $z_a\lesssim 1/k_{\rm F}\approx 0.5 \mathrm{\AA}$, which lies beyond the applicability of our theory.

The previous description introduces several scales into our system which are useful for characterizing its behavior (some of them are also featured
by a spatially local description, such as the Drude model, see 
below). We have the reduced plasma wavelength $\lambdabar_{\rm p}=c/\omega_{\rm p}$ and the diffusion constant $D=\Gamma \lambdabar_{\rm p}^{2}$ related, respectively, with the plasma vibrations and the diffusion of the electromagnetic field in the metal. The latter can also be connected to the penetration of the electromagnetic radiation in the metal which is described by the skin-depth length $\delta_{a}=\sqrt{2D/\omega_a}$, evaluated at the atomic transition frequency \cite{Jackson75}.
Additionally, our spatially nonlocal description is characterized by the electron's mean-free path $\ell=v_{\rm F}/\Gamma$ and the Thomas-Fermi wavelength $\lambda_{\rm TF}=v_{\rm F}/(\omega_{\rm p}\sqrt{3})$, which describe the ballistic behavior of the electrons between two collisions and screening effects in the plasma, respectively \cite{Manfredi05}.

While we can safely assume that the distribution describing the electronic fluid obeys the zero-temperature Fermi-Dirac distribution for a wide range of temperatures ($0\leq T\ll T_{\rm F}$), the expression for $\Gamma$ can feature a rather different behavior depending on both the temperature and the nature of the collisions. For instance, if we consider an (infinite) bulk made of ideal metal, where all the lattice atoms are perfectly periodically aligned and no defects are present (perfect crystal), dissipation predominantly arises from scattering processes between the elementary particles in the system. 
In this case, within a quantum description, the Bloch-Gr\"{u}neisen formula predicts that, at low temperature, $\Gamma(T)\propto T^m$
where $m\geq2$ (e.g. $m=2$ for electron-electron scattering and $m=5$ for electron-phonon scattering) \cite{Bloch29,Bloch30,Bass90}. For realistic systems which feature deviations from this ideal configuration, destroying the periodicity of the crystal (e.g. due to impurities in the crystal as well as other static defects, including dislocations and the sample's surface), a residual resistivity appears at $T=0$ \cite{Bass90}.

\subsection{Limiting behaviors}

The Boltzmann-Mermin permittivites feature two limiting regimes. For values $|u|\rightarrow\infty$, obtained when $v_{\rm F}\ll\omega/k$ and/or $k\ell\ll 1$, the dielectric functions in Eqs. \eqref{permittivities} recover the usual local Drude description of the metal \cite{Jackson75}, i.e. $\epsilon_l(\omega,k)=\epsilon_t(\omega,k) \to\epsilon(\omega)=1-\op^2\left[\omega(\omega+\imath\Gamma)\right]^{-1}$. Physically, these values correspond to a regime where the phase velocity ($v_{\rm ph}=\omega/k$) of the electromagnetic radiation is much larger than the Fermi velocity and/or the field's wavelength is much larger than the electron's mean free path. 
The opposite limit, $|u|\rightarrow 0$, is connected to wavelengths that are sufficiently small so that they resolve the electron's ballistic motion, i.e. $v_{\rm F}\gg v_{\rm ph}$ and/or $k\ell \gg 1$ \cite{Reiche17}. In this case, the permittivities take the form
\begin{subequations}
\label{limit}
\begin{align}
	\epsilon_l\left(\omega,k\right)
		&\approx 1+\frac{3}{k^2\lambda^{2}_{\rm TF}}
			+\imath\frac{\omega \Gamma^{\rm L}_{l}(k)}{\omega_{\rm p}^{2}},\\
	\epsilon_t\left(\omega,k\right)
		&\approx 1-\frac{3}{k^2\lambda^{2}_{\rm TF}}
			+\imath\frac{\omega_{\rm p}^2}{\omega\Gamma^{\rm L}_{t}(k)},
\end{align}
\end{subequations}
where $\Gamma^{\rm L}_{l}(k)=3\pi\omega_{\rm p}^{4}/(2v_{\rm F}^{3}k^{3})$ and $\Gamma^{\rm L}_{t}(k)=4v_{\rm F}k/(3\pi)$ are the semiclassical Landau damping rates for longitudinal and transverse fields propagating in a fermionic plasma \cite{Landau46,Van-Kampen57}.
Landau damping can be understood as resulting from a net energy-momentum transfer from the field to the electronic fluid. It occurs both in classical and quantum gases when the phase velocity of the radiation is smaller than the speed of the quasi-particles. Landau damping is therefore closely related to the quasi-particle's distribution in phase space. This means that in our case the expressions for $\Gamma^{\rm L}_{l,t}(k)$ are directly connected to the specific form of the equilibrium distribution $f_{0}$ that we chose at the beginning of our analysis.
It is worth pointing out that in Eqs. \eqref{limit} the collision rate $\Gamma$ completely dropped out from the expressions for the permittivity, showing that Landau damping prevails as the main source of attenuation for the field (see also Ref. \cite{Svetovoy05}).

The above-described regimes and the corresponding length scales are inherited by the surface impedances through Eqs. \eqref{surfaceimpedance}, where the role of the three-dimensional wave vector is roughly taken by its in-plane component $p$ [see also Eq. \eqref{SI2} for a more detailed discussion of the behavior of $Z^{\rm TE}(\omega, p)$].
This means that Eqs. \eqref{surfaceimpedance} tend to Eqs. \eqref{SurfImpadance} if $p\ell \ll 1$ and, as we are going to see below, the Casimir-Polder interaction recovers the local behavior in this limit. 
Conversely, for $p\ell \gg 1$, Landau damping will act as the main source of dissipation in the reflection coefficients of Eqs. \eqref{reflectioncoefficients}, modifying the interaction accordingly.

\subsection{Free energy}
\begin{figure}
  \centering
    \includegraphics[width=0.45\textwidth]{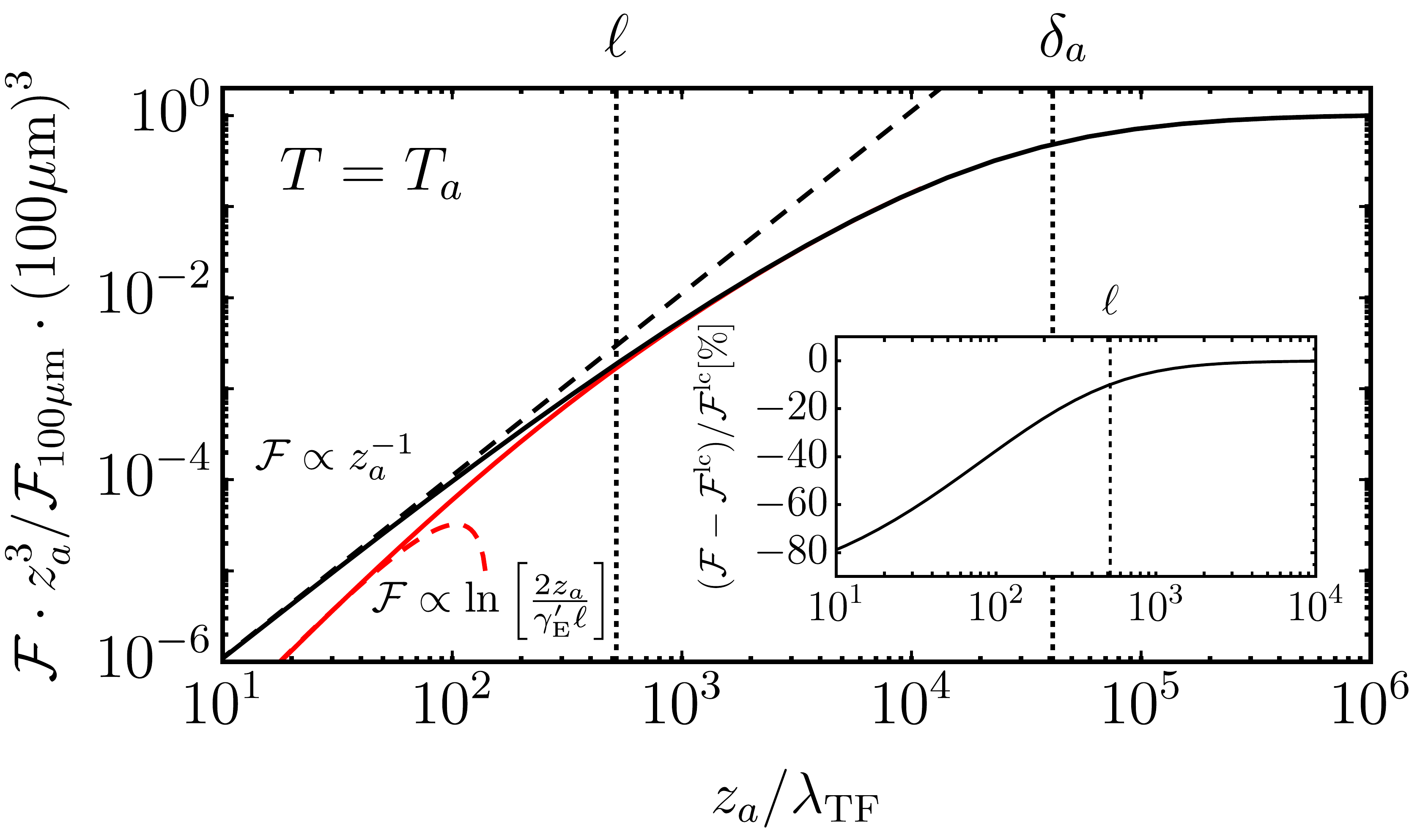}
    \caption{(Color online) Magnetic Casimir-Polder free energy $\mathcal{F}$ 
    as a function of the atom-surface separation $z_a$. 
    We multiply the curve by $z_a^3$ and normalize 
    to the corresponding local value at $z_a=100\mu$m. 
    The magnetic dipole is oriented 
    parallel to the surface (anisotropic polarizability $
    \beta_{xx}^T=\beta_{yy}^T$, $\beta_{zz}^T=0$). 
    The transition frequency is $\omega_a=2\mu$eV and 
    the atom is trapped above a gold surface ($
    \omega_{\rm p}=9$eV, $\Gamma=\Gamma_{\rm Au}=30$meV,
     $\vf\approx\alpha_{\rm fs}c
    $ with $\alpha_{\rm fs}$ the fine-structure constant). 
    The system is in equilibrium at temperature $T=T_a$. 
    The free energy obtained using a local Drude model $
    \mathcal{F}^{\rm lc}$ (black line) changes its power 
    law dependence $z_{a}^{-3}\to z_{a}^{-1}$ for separations 
    much smaller than the skin-depth $\delta_{a}$ 
    (see Eq. (\ref{Flocalsmall}) and 
    discussion below). 
    The nonlocal description (red line) deviates from the previous curve
    for distances shorter than the electron mean-free 
    path $\ell=v_{\rm F}/\Gamma$.
    At short separations, the distance dependence is 
    no longer described by a power law
    but is logarithmic according to the expression 
    in Eq. \eqref{Fnonlocalsmall} (dashed, red).
    In the inset we give the 
    relative deviation of the nonlocal from the local material model.  
    \label{FreeEn}}
\end{figure}

In order to illustrate the impact of spatial dispersion on the atom-surface interaction, we analyze in the following the behavior of the magnetic Casimir-Polder free energy.
The nonlocal material model and its local limit differ in their dependence on the radiation's wave vector. The exponential in the Green tensor of Eq. (\ref{RIGreen}) selects the wave vectors dominating the interaction as $p\sim1/z_a$. Hence, the impact of spatial dispersion on the Casimir-Polder free energy can be seen best as a function of atom-surface separation.
After a rotation in the complex frequency plane \cite{Dzyaloshinskii61}, the magnetic Casimir-Polder free energy in Eq. \eqref{FreeEnergy} can be written as
\begin{align}
	\mathcal{F}(z_{a},T)=\pi k_{\rm B}T \sideset{}{'}\sum_{n=0}^{\infty}
					\Delta^{T}(z_{a},\imath n \nu),
\label{MatsubaraFreeEn}
\end{align}
where $\nu=2\pi k_{B}T/\hbar$ is the first Matsubara frequency and the prime indicates that the first term of the sum has to be taken with a prefactor $1/2$. Here, we have defined the function 
\begin{equation}
	\Delta^{T}(z_a,\omega)=-\frac{1}{\pi}
							\text{Tr}\left[\underline{\beta}^T(\omega)
							\cdot\underline{h}^{T}(z_a,\omega)\right].
\label{DeltaMCP}
\end{equation}
The magnetic polarizability constrains the contributing frequencies to $\omega_a$ and sets the maximally relevant length scale to the wavelength $\lambda_{a}=2\pi c/\omega_{a}$ (in the centimeter range for magnetic interactions). In the following, we focus on the limit $z_{a}\ll \lambda_{a}$ of $\mathcal{F}(z_{a},T)$, where retardation effects can safely be neglected. 
In Fig. \ref{FreeEn}, we report the results obtained for $\mathcal{F}(z_{a},T)$ using a purely local (Drude) dielectric model for the metallic surface and the nonlocal expressions in Eqs. \eqref{permittivities}. 
We notice that the two descriptions lead to identical results at large distances. In particular, for separations larger than the metallic skin-depth $\delta_{a}$, the free energy scales as $z_a^{-3}$ in both cases.
For $z_{a}\ll \delta_{a}$, however, the local description yields a free energy which behaves as 
\begin{align}\label{Flocalsmall}
	\mathcal{F}^{\rm lc}
		&\approx \pi k_{\rm B}T
				 \sideset{}{'}\sum_{n=0}^{\infty}
				 \frac{n^{2}\nu^{2}}{z_a c^{2}}
				 [\epsilon(\imath n\nu)-1]\Phi^{T}(\imath n\nu) \nonumber\\
		&\approx \frac{\pi k_{\rm B}T}{\lambdabar_{\rm p}^2 z_a}
				 \sideset{}{'}\sum_{n=0}^{\infty}
				 \frac{n}{n+\frac{\Gamma}{\nu}}\Phi^{T}(\imath n\nu),
\end{align}
where we define
\begin{equation}
	\Phi^{T}(\omega)=\mu_{0}
					 \frac{\text{Tr}\left[\underline{\beta}^T(\omega)\right]
					 	   +\beta_{zz}^T(\omega)}
					 	  {(8\pi)^{2}}.
\label{phi}
\end{equation}
The expression in Eq. \eqref{Flocalsmall} is obtained using the non-retarded expression for the Green tensor in Eq. \eqref{RIGreen}, which is formally equivalent to taking the limit $c\to \infty$. The permittivity is significantly different from its vacuum value for $\omega<\omega_{\rm p}$ and, since the polarizability is different from zero only for $\omega\lesssim \omega_{a}\ll \omega_{\rm p}$, we can consider the limit $\omega/(c p)\ll1$ and use the approximation
\begin{equation}
	r^{\rm TE}(\omega,p)\approx \frac{\epsilon(\omega)-1}{4}
								\frac{\omega^{2}}{c^{2}p^{2}}.
\label{apprTE}
\end{equation}
The contribution of the TM-polarization gives only minor corrections in the limit described above.

The free energy for the nonlocal description starts to differ from the local one for distances smaller than the mean-free path $\ell$. Since the relevant frequencies are such that $\omega<\omega_{a}\ll \Gamma$, in the limit $z_{a}\ll \ell$ we can use the result of Appendix \ref{appendixentropylow} and write
\begin{align}\label{Fnonlocalsmall}
	\mathcal{F}^{\rm nl}
			&\approx -\pi k_{\rm B}T
					 \ln\left[\frac{2z_a}{\gamma'_{\rm E}\ell}\right]
					 \sideset{}{'}\sum_{n=0}^{\mathrm{Int}[\Gamma/\nu]}
					 \frac{4n\nu}{v_{\rm F}\lambdabar_{\rm p}^{2}}
			         \Phi^{T}(\imath n\nu) ,
\end{align}
where $\gamma'_{\rm E}=e^{-\gamma_{\rm E}}$ with $\gamma_{\rm E}$ the Euler-Mascheroni constant and ${\rm Int}[x]$ gives the largest integer smaller than $x$. In accordance with our approximations, we also set a limit to the number of Matsubara frequencies involved in the sum. 
Notice that including nonlocality in the description of the metal reduces the strength of the interaction \cite{Esquivel03}, also changing its distance dependence.  
With respect to the local case, spatial dispersion quickly induces an attenuation larger than $10\%$ as soon as the atom-surface separation is shorter than the mean-free path (see inset of Fig. \ref{FreeEn}). It is also interesting to remark that this correction would not occur in simpler spatially nonlocal material models. For instance, in the hydrodynamic description (see for example \cite{Moeferdt18}), one models the quasi-particles of the electronic plasma by means of the hydrodynamic Euler equation rather than the Boltzmann equation, i.e. as a compressible fluid of particles that interact via collisions and that are repelling each other via the Fermi pressure.
It turns out that only the longitudinal permittivity is modified as a function of the Fermi-velocity \cite{Esquivel03}. The transverse permittivity, however, equals the (local) Drude description, leaving $r^{\rm TE}$ unchanged. Since the latter provides the dominant contribution for separations $z_a\ll\ell$, spatial nonlocality induced by the hydrodynamic model has a negligible impact on the magnetic Casimir-Polder free energy.

\section{Casimir Entropy}
\label{EntropyPhenomenology}

In order to set the context for the analysis provided in the subsequent sections, we need to briefly review the general characteristics of the Casimir entropy. We focus first on features that neither depend on the specific geometry nor on the specific material properties of the system. For clarity, however, we consider later the magnetic Casimir-Polder interaction as a specific example.

\subsection{General framework}

As was pointed out in previous works (see for example \cite{Intravaia12b}), the first important remark is that all the thermodynamical properties related to the Casimir effect (entropy, free energy, etc.) are differential quantities. They focus on the  interaction and arise from the difference of the same quantity calculated for two different configurations involving interacting versus non-interacting bodies. Usually, the parameter distinguishing these two configurations is the separation between the objects, which is finite for the interacting objects and infinity for the non-interacting ones.
The Casimir entropy is derived from the negative derivative with respect to temperature of the Casimir free energy and it can always be written as
\begin{align}
	\mathcal{S}(z_a,T)=
		-\frac{d}{dT}\sideset{}{'}\sum_{n=0}^{\infty}
		\pi k_{\rm B}T\Delta^{T}(z_{a},\imath n \nu).
\label{entropy}
\end{align}
The details of the function $\Delta^{T}(z_a,\omega)$ depend on the geometry and on the properties of the system (see Eq. \eqref{DeltaMCP} for the magnetic Casimir-Polder interaction). In general, $\Delta^{T}(z_a,\omega)$
is analytic in the upper part of the complex $\omega$-plane and it allows for an intrinsic temperature dependence which stems from the physical properties of the bodies involved in the interaction.  Given that $[\Delta^{T}(z_a,\omega)]^{*}=\Delta^{T}(z_a,-\omega^{*})$ (crossing relation \cite{Ford88}), $\Delta^{T}(z_a,\omega)$ is a real-valued function along the positive imaginary frequency axis, i.e. for $\omega=\imath \xi$ ($\xi>0$).
The function $\varrho^{T}(z_a,\omega)=-\partial_{\omega}\mathrm{Im}[\Delta^{T}(z_{a},\omega)]$ gives the system's \emph{differential} mode density, i.e. the difference between the number of modes with a frequency in $[\omega,\omega+d\omega]$ for the interacting configuration and the same number for the non-interacting configuration. Since $\Delta^{T}(z_a,\omega)$ vanishes for $|\omega|\to \infty$ \cite{Note0} and from the crossing relation it follows that $\mathrm{Im}[\Delta^{T}(z_{a},0)]=0$, the integral of $\varrho^{T}(z_a,\omega)$ over all positive frequencies is zero. This indicates that the total number of modes (which can be infinite in both cases) does not change when going from the interacting to the non-interacting configuration. Their spectral distribution, however, does.

From Eq. \eqref{entropy} we obtain that at high temperature the entropy tends to the value 
\begin{equation}\label{Slarge}
	\mathcal{S}(z_a,T)\stackrel{T\to\infty}{\approx}
		-\pi k_{\rm B}
		\frac{\Delta^{T}(z_{a},0)}{2}
		-\pi\frac{k_{\rm B}}{2}\partial_{T}\Delta^{T}(z_{a},0).
\end{equation}
The expression of $\mathcal{S}(z_a,T)$ for $T\to 0$ is somewhat more involved. Following the approach described in Appendix \ref{Entropyanalysis} (see also Ref. \cite{Intravaia08}), we obtain
\begin{align}\label{lowTemp}
	\mathcal{S}(z_a,T)
		&\stackrel{T\to0}{\approx}\pi k_{\rm B}
			\frac{\Delta^{T}
			\left(z_{a},\imath\nu\right)
			-\Delta^{T}(z_{a},0)}{2} 
		\\\nonumber
		&\quad
			-\frac{2}{3}
			\frac{\pi^{2} k_{B}^{2}}{\hbar}  
			\partial_{\nu}\Delta^{T}
			\left(z_{a},\imath\nu \right)T
		\\\nonumber
		&\quad
			-\frac{\hbar}{2}
			\int_{0}^{\infty}\text{d}\xi\partial_{T}\Delta^{T}(z_{a},\imath \xi).
\end{align}
In the following, we focus on the fist two terms of the above expression, since it turns out that the last term can be safely neglected for most of our considerations (see Appendix \ref{Entropyanalysis}). 
The two previous equations link the behavior of the Casimir entropy at low and high temperature to the behavior of the function $\Delta^{T}\left(z_{a},\omega\right)$ and its derivative within a region of the complex plane around $\omega\sim 0$. Since low-frequency effects are strongly influenced by dissipative mechanisms \cite{Reiche17}, both extremal cases described in the Eqs. (\ref{Slarge}) and (\ref{lowTemp}) are closely intertwined with dissipation in the system.
Furthermore, the limit of high temperatures for the Casimir entropy can be connected to the limit of large separations for the Casimir free energy (see for example Ref. \cite{Milton17a}). The latter implies that the Casimir entropy is positive for $T\to \infty$, if the interaction leads to an attractive force at large separations. By the same token, the Casimir entropy at high temperature is negative, if the Casimir force is repulsive for large distances.

The situation for $T\to 0$ is considerably more complicated.  
If $\Delta^{T}\left(z_{a},\omega\right)$ is well-behaved for $\omega\sim 0$,
Eq. \eqref{lowTemp} can be indeed further simplified to read (see also Appendix \ref{Entropyanalysis})
\begin{align}
	\mathcal{S}(z_a,T)
		&\approx \frac{\pi^{2}}{3} 
			  	 \frac{k_{\rm B}^{2}}{\hbar}\varrho^{0}(z_a,0)T.
\label{approxEntropy}
\end{align}
We find that in these cases the entropy vanishes  at least linearly with the temperature. 
Notice that $\varrho^{0}(z_a,0)\not=0$ for any ohmic material but it may vanish otherwise.
For the systems considered in the literature, calculations have shown that $\varrho^{0}(z_a,0)$, when it is nonzero, is positive for an interaction being repulsive at large distances while it is negative for an attractive interaction (similar arguments can be used for Eq. \eqref{lowTemp}). 
This implies that $\mathcal{S}(z_a,T)$ can be a non-monotonic function of $T$ vanishing at least in one intermediate temperature \cite{Bezerra02,Canaguier-Durand10a,Milton15a,Milton17a,Note2}. 
For an analysis of the entropy's low-temperature behavior and its interplay with the geometry of the system and the material properties see also Ref. \cite{Ingold15}.

A non-vanishing entropy, however, can occur when 
\begin{equation}
	\lim_{T\to 0}\Delta^{T}(z_{a},\alpha T)
		\not=\lim_{T\to 0}\lim_{\omega\to 0}\Delta^{T}(z_{a},\omega), 
		\quad \forall \alpha\not=0.
\label{nonCommLimit}
\end{equation}
When the limits do not commute, $\mathcal{S}(z_a,T)$ approaches for $T\to 0$ the constant value 
\begin{equation}
	\mathcal{S}_{0}= \lim_{T\to 0} \pi k_{\rm B}
					 \frac{\Delta^{T}\left(z_{a},\imath\nu\right)
					 	   -\Delta^{T}\left(z_{a},0\right)}{2}
\label{general-Entropydefect}
\end{equation}
rather than going to zero.
Once again, the sign of this constant depends on the configuration and while for attractive interactions it is negative \cite{Bezerra04,Intravaia08,Intravaia09}, for repulsive forces it is positive \cite{Haakh09b}.

In the next sections, we analyze some mechanisms that can lead to Eqs. \eqref{approxEntropy} and \eqref{nonCommLimit}. 
In particular, 
we will establish a connection with the thermodynamic 
properties of overdamped resonances (eddy currents), with special focus on the case of the magnetic Casimir-Polder interaction.

\subsection{Entropy for the magnetic Casimir-Polder interaction}

\begin{figure}
  \centering
   \includegraphics[width=0.48\textwidth]{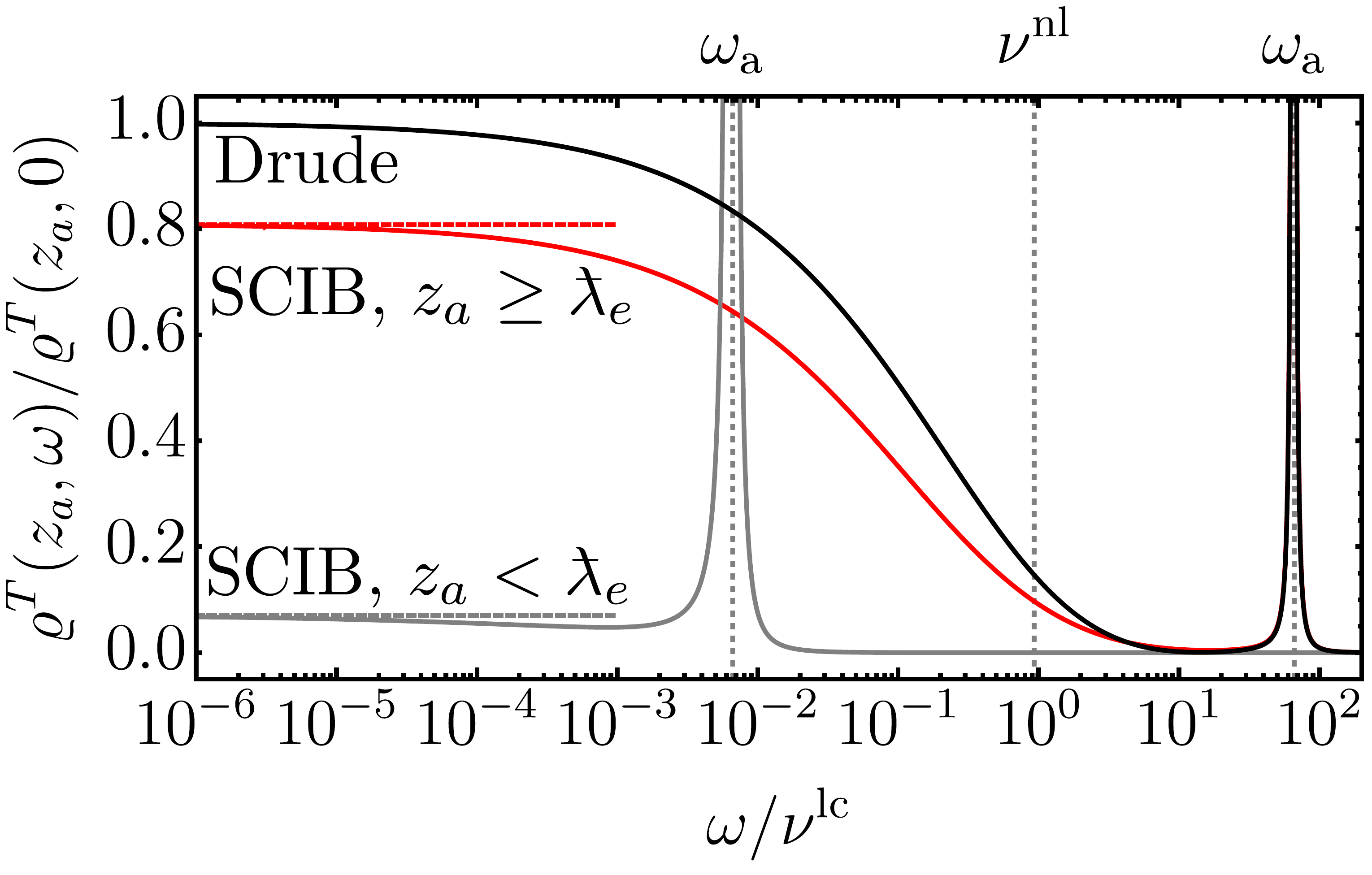}
    \caption{(Color online) Numerical evaluation of the Casimir-Polder mode density 
    for frequencies $\omega \lesssim\nu^{\rm lc}=D/z_a^2$ at temperature $T=T_a$ 
    normalized to the local (Drude) limit $\varrho^{T}(z_a,0)$ [see Eq. 
    (\ref{lowfrequencies})]. As in Fig. \ref{FreeEn}, we plot the situation for an 
    anisotropic dipole with transition frequency $\omega_a=2\pi\times 480$ MHz and 
    report the results using the local (Drude) material model 
    (black, solid line) 
    and the nonlocal SCIB model (red, solid line).
    In both scenarios, the (differential) mode density exhibits 
    low-frequency quasi-
    normal excitations which, in part, can be traced back to the eddy modes of the 
    configuration. For frequencies $\omega\ll\nu^{\rm lc}$, the mode densities 
    approach their respective local (normalization) and nonlocal (red, dashed) 
    limits given by Eqs. (\ref{lowfrequencies}) and (\ref{nonlocalmodedensity}). 
    However, at frequencies near the atomic resonance $\omega\sim\omega_a$, the 
    different models coincide.
    We choose $z_a=100~\lambdabar_{\rm p}$ and $\Gamma=0.01~\Gamma_{\rm Au}$ 
    in order to ensure that $z_a\geq\lambdabar_{\rm e}$ [see Eq. \eqref{xinl}] 
    and the (nonlocal) eddy modes can have a 
    significant impact according to Eq. (\ref{lambdae}). 
    Also, for the parameters used here, the typical 
    excitation energies of eddy modes in both material models lay within the same 
    range, i.e. $\nu^{\rm lc}\sim\nu^{\rm nl}$. For comparison, we display the 
    result regarding separations $\tilde{z}_a=\lambdabar_{\rm p}$ smaller than the 
    cutoff wavelength $\lambdabar_{\rm e}$ (gray, solid line). Again, the mode density 
    approaches its nonlocal limit for $\omega\to 0$ (gray, dashed), but its impact, 
    with respect to the (shifted) atomic transition frequency ($\nu^{\rm lc}=D/
    z_a^2\rightarrow D/\tilde{z}_a^2$) and the local material model, is 
    significantly reduced [see discussion below Eq. (\ref{lambdae})]. \label{ModeDensity}}
\end{figure}

As an example of the previous results, let us consider first the case of the magnetic Casimir-Polder interaction, where the metallic half-space is described by the Drude model with a constant damping $\Gamma$.
Using the definition in Eq. \eqref{phi} and the result in Appendix \ref{appendixentropylow}, we obtain 
\begin{align}
	\varrho^{T}(z_{a},0)\approx 		
		\frac{\Phi^{T}(0)}{z_{a}D}>0.
\label{lowfrequencies}
\end{align}
For the nonlocal SCIB model the mode density at low frequencies additionally depends on the ratio of the atom-surface separation and the electron's mean free path $\ell$. While for $z_a\gg\ell$ we recover Eq. \eqref{lowfrequencies}, in the opposite limit we obtain (see Appendix \ref{appendixentropylow})
\begin{align}
\label{nonlocalmodedensity}
	\varrho^T(z_a,0)
		&\overset{z_a\ll \ell}{\propto} -4
			\frac{\Phi^{T}(0)}{\lambdabar_{\rm p}^2v_{\rm F}}
			\ln\left[\frac{2z_a}{\ell}\right]>0.
\end{align}
In both cases, according to Eq. \eqref{approxEntropy}, $\mathcal{S}(z_a,T)$ vanishes linearly with the temperature.
The entropy approaches zero from the positive side, matching the above discussion (see also Fig. \ref{FreeEn} and Ref. \cite{Haakh09b}).
The numerical evaluation of the mode density for low frequencies using the local and the nonlocal material model can be found in Fig. \ref{ModeDensity}. The result confirms our prediction and we see that, in general, the inclusion of nonlocality leads to a reduction of the mode density with respect to the local limit. This behavior is also agreement with the reduction of the magnetic Casimir-Polder interaction observed in Fig. \ref{FreeEn}.

\begin{figure}
  \centering
    \includegraphics[width=0.48\textwidth]{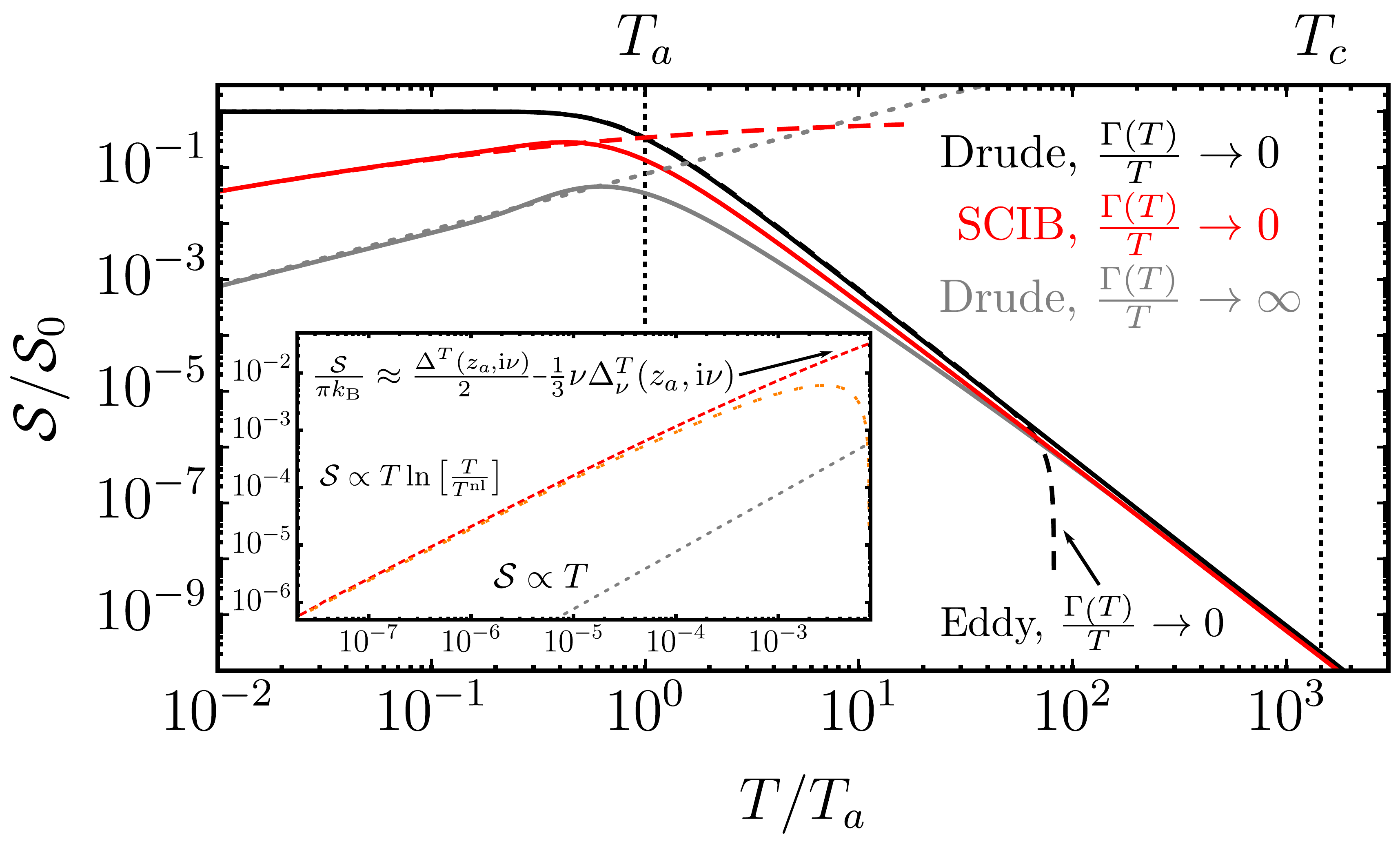}
    \caption{(Color online) Magnetic Casimir-Polder entropy in the low-temperature 
    regime normalized to the value $\mathcal{S}_{0}$ defined in Eq. 
    \eqref{entropydefect}. 
    The numerical results for an anisotropic dipole-configuration (dipole moment 
    parallel to the $xy$-plane) modeled as a two-level system with transition 
    frequency $\omega_a=2\mu$eV are reported. 
    The atom is placed at $z_a=1\mu$m separation from the metallic bulk.
    For comparison, we plot the entropy using a local Drude model with constant 
    dissipation rate (gray, solid line) vanishing linearly for low temperatures 
    as given by Eq. \eqref{approxEntropy}  (gray dashed).
    The local Drude model with a temperature-dependent dissipation rate $\Gamma(T)=
    \Gamma_{\rm Au}[T/T_c]^5$ (electron-phonon scattering), 
    instead, goes to a constant for $T\to 0$ (black, solid line) 
    as well as the contribution solely stemming from Eddy modes (black, dashed).
    Using the nonlocal material model (red, solid line), the entropy 
    does vanish for $T\to 0$ also if $\Gamma(T)/T\to 0$. The 
    asymptotic curve of the nonlocal entropy (red, dashed) is obtained from Eq. 
    \eqref{lowTemp}. The inset shows the details of the behavior 
    for $T\to 0$ [local (gray) and nonlocal (orange)] in connection with the 
    expression in Eq. \eqref{lowTemp}.
     \label{fig:S}}
\end{figure}

Next, let us consider the case of a metal where the dissipation rate is assumed to follow the behavior of a perfect crystal, i.e. vanishing for $T\to 0$ faster than linearly.
We note that in this case 
the function $\Delta^{T}(z_a,\omega)$ features the discontinuos behavior described in Eq. \eqref{nonCommLimit}. Due to the vanishing diffusion coefficient $D(T)=\Gamma(T) \lambdabar_{\rm p}^{2}$, the divergence of the expression given in Eq. \eqref{lowfrequencies} prevents the use of Eq. \eqref{approxEntropy}. Starting from Eq. \eqref{lowTemp}, the second term tends to zero leaving, as expected, a residual Casimir entropy with the value 
\begin{align}
\label{entropydefect}
	\mathcal{S}_{0}
		=4\pi\kb
		\frac{\Phi^{0}(0)}{z_a^3}F
			 \left(\frac{z_a}{\lambdabar_{\rm p}}\right).
\end{align}
In the previous equation we define
\begin{equation}
	\quad F(x)=-\int_{0}^{\infty}{\rm d}yy^2e^{-2y}
			   \frac{y-\sqrt{y^2+x^2}}{y+\sqrt{y^2+x^2}}
\end{equation}
which is a positive, monotonically increasing function (see Fig. \ref{fig:S}). For $z_{a}\ll \lambdabar_{\rm p}$, $F\to 1/4$ recovering the value already reported in Ref. \cite{Haakh09b}.

If we use the nonlocal description of the metal instead [see Eqs. \eqref{permittivities}], the entropy at zero temperature vanishes even if $\Gamma(T)/T \to 0$ for $T\to0$.
From Eq. \eqref{lowTemp}, using Eq. \eqref{final2} and that $\lim_{T\to 0}\Delta^{T}(0)=0$, the expression for the low-temperature entropy reduces to
\begin{align}\label{Slow1}
	\frac{\mathcal{S}(T\to 0)}{\pi k_{\rm B}}
		&\approx-\frac{2}{9}\nu
				\frac{\Phi^{T}(0)}{v_{\rm F}\lambdabar_{\rm p}^{2}}
				\ln\left[\left(\frac{2}{\gamma'_{\rm E}}\right)^{3}
				\frac{\nu}{\nu^{\rm nl}}\right]
		\\\nonumber
		&
		\propto 
		-T
		\ln
		\left[
		\frac{T}{T^{\rm nl}}
		\right],
\end{align} 
where the frequency scale $\nu^{\rm nl}$ is set to
\begin{equation}
 \nu^{\rm nl}=\frac{4 v_{\rm F}}{3\pi z_{a}}
		\frac{\lambdabar_{\rm p}^{2}}{z_{a}^{2}} = 
		\frac{\lambdabar_{\rm p}^{2}}{z_{a}^{2}}\Gamma^{\rm L}_{t}\left(1/z_{a}\right)
\label{nonlocalfrequencyscale}
\end{equation}
and $T^{\rm nl}$ gives a corresponding temperature.
The previous expressions show that, in the nonlocal case, $\mathcal{S}(T\to 0)$ vanishes following a non-algebraic relation highlighting the important role that Landau damping plays in determining the entropy of the system at low temperature. 

In Fig. \ref{fig:S} we depict the non-monotonic behavior of the magnetic Casimir-Polder entropy for our specific system.
We present the numerical results considering both constant and temperature-dependent dissipation rates and employ the nonlocal material description and the local approximation. As predicted, independent from the behavior of the dissipation rate, the entropy of the nonlocal material model vanishes for $T\to 0$: It behaves linearly ($\propto T$) for constant dissipation and goes as $\mathcal{S}\propto -T\ln \left[T/T^{\rm nl}\right]$ when $\Gamma(T)/T\to 0$. 
Conversely, for the local description, $\mathcal{S}$ goes to zero for constant $\Gamma$, but approaches the value $\mathcal{S}\to\mathcal{S}_0$, if for example $\Gamma \propto T^{5}$ \cite{Haakh09b}.
Note that the transition from what we call ``high temperatures'' to ``low temperatures'' is marked in terms of the atomic transition temperature $T_a$. In addition, for temperatures $T\gg T_c\equiv\hbar v_{\rm F}/(k_{\rm B} z_a)$ ($\sim16$K for $z_a=1\mu$m and $v_{\rm F}=\alpha_{\rm fs}c$) spatial nonlocality can be, to very good approximation, neglected and the curves of the different scenarios become indistinguishable.

\subsection{Nernst Theorem and the plasma-Drude controversy}\label{Sec:PDControversy}

The previous results are relevant with respect to one of the most controversial and still discussed laws of thermodynamics, i.e. the Nernst theorem \cite{Morse62,Kittel80} (see Refs. \cite{Belgiorno03,Masanes17,Shirai18} and references therin for some recent discussions and Ref. \cite{Kox06} for a historical perspective).
In one of its formulations, due to Planck in 1911, this thermodynamic principle implies that the entropy of a system at equilibrium goes to zero at $T=0$ \cite{Fermi56}. This finds its justification within quantum theory and the often found uniqueness of the (well-ordered) ground state of systems. However, many systems have challenged this formulation featuring different non-zero residual entropies as a symptom of some level of disorder which persists when the temperature is extrapolated to zero. 
Nernst himself gave a version of his law stating that a finite size system at equilibrium has an entropy at zero temperature, which is independent of any external macroscopic parameters (e.g. volume, pressure, temperature). 
For infinite systems, some investigations \cite{Casimir63,Fisher64,Griffiths65,Aizenman81} have pointed out that the entropy may be crucially related to the distribution of low-energy excitations in the limit $T\to 0$. The correct value is then related to a specific thermodynamic limiting procedure, which amounts to dividing the free energy by some of the system's extensive parameters \cite{Fisher64,Griffiths65,Huang03}. This would, e.g., involve the existence of a bulk thermodynamic entropy per particle \cite{Leff70}. A connection to the degeneracy of the ground state can be re-established when the latter is suitably interpreted \cite{Masanes17,Aizenman81}. 

Although one can construct models which violate Nernst theorem \cite{Leff70,Bonner62}, the physical origin of a finite distance-dependent interaction entropy in Eq. \eqref{entropydefect} remains puzzling, especially when contrasted with the nonlocal calculation which does not pose any issue. 
From the standpoint of condensed matter theory, the Drude model is certainly an oversimplification of the physics occurring in a conductor. 
In particular, for $\Gamma\to 0$, the mean-free path $\ell$ becomes increasingly larger indicating that a metal at low frequencies is inadequately described by the local description. 
Using the Drude model artificially keeps the metal in the local regime, even in the limit $\Gamma\to 0$ (see also Eq. \eqref{SI2} below). 
Conversely, the SCIB model [Eqs. \eqref{permittivities}] approaches its nonlocal limit and the scattering-induced damping as well as its temperature dependence loose their relevance for the material's bulk response. Since at low temperature the Casimir entropy vanishes for the nonlocal model, one might be then tempted to conclude that nonlocality is directly responsible for the vanishing entropy. We are going to see in the following, however, that this is not the case and that the actual reasons lie deeper in the statistical properties of the electron gas.

Given the general structure of the relations in Eqs. \eqref{nonCommLimit} and \eqref{general-Entropydefect} it is not surprising that a nonzero entropy was found in several other configurations.
We have already mentioned that results similar to Eq. \eqref{entropydefect} have been reported for the plane-plane Casimir interaction (further examples can be found in Refs. \cite{klimchitskaya08,bordag10}), where a finite \emph{negative} entropy was found \cite{Bezerra04}. The nonlocal description of the metal in Eqs. \eqref{permittivities} led instead to a vanishing entropy \cite{Svetovoy05,svetovoy08}. 
Based on this result, the authors of \cite{Bezerra04} concluded that the Nernst theorem is not fulfilled by the Drude model, claiming its thermodynamic inconsistency.
The distrust in the Drude description can be understood within a broader context involving room temperature experimental results for the Casimir force between a gold coated sphere and a plane: The measurements performed in a range of distances $\lesssim 1 \,\mu$m  \cite {Decca05,Decca07,Banishev13,Bimonte16,Liu19a} are in disagreement with the prediction obtained using the Lifshitz formula \cite{Lifshitz56} and the Drude model. They agree, however, with a theoretical description where no further changes are made beside setting the dissipation rate to zero, the so-called plasma model. 
The Drude model, however, appears to better agree with measurements in a range of distances between $\sim 1 \, \mu$m and $\sim 7 \, \mu$m, once a residual background electrostatic component of the force has been removed using a two-parameter fitting procedure \cite{Sushkov11}.
In combination with the experimental results, the non-vanishing entropy at zero temperature has been used to criticize the Drude model in favor of the non-dissipative plasma description for which $\mathcal{S}(T\to0)\to 0$ \cite{Bezerra04}. 
Also, as discussed above, within the Lifshitz description of the Casimir interaction, the behavior of the entropy in the limit $T\to 0$ and of the force at large distances and/or temperature originate from the low frequency response of the material, where Drude and plasma model notably differ. Experimentally, the measurement of the dielectric function for low frequencies is a non-trivial task and results can differ significantly from sample to sample \cite{svetovoy08a}. Since the dielectric function can be measured in a finite frequency interval only, one is obliged to apply extrapolation techniques \cite{reyes18}. 
It is important to mention that, despite the entropy vanishes at zero temperature, the spatially nonlocal permittivity functions in Eqs. \eqref{permittivities} and the SCIB model \cite{Svetovoy05} do not resolve the issue on the quantitative description of the force measurements at room temperature, since the prediction is very similar to that of the Drude model. 
A solution for this problem might require a more careful modeling of the material's optical response as well as a deeper understanding of the underlying physics.

To the best of our knowledge, a precise and comprehensive understanding of the physical origins for the different outcomes of the zero-temperature entropy depending on the chosen material model is missing so far, even from the theoretical point of view.  In the following, rather than directly focusing on the  Nernst theorem, we investigate the thermodynamics of the underlying processes.
To this end, we will connect the zero-temperature entropy to the statistical properties of the eddy modes and provide the physical foundation for explaining the different behaviors.


\section{Overdamped modes in local materials}
\label{LocalEddyModes}
In general, eddy currents are diffusive bulk modes
that occur in conducting materials under the influence of slowly varying magnetic fields \cite{Maxwell71,Jackson75,Henkel10,Henkel19b}.
A physical interpretation of the peculiar behavior of the Casimir entropy in the plane-plane configuration and for the Drude model has been provided 
in Ref. \cite{Intravaia09} in terms of eddy or Foucault current modes (see also Refs. \cite{Torgerson04,Bimonte07a,Svetovoy07}). 
Some of the properties of these currents are not a unique feature of the plane-plane configuration.
In general, eddy currents are diffusive modes
that occur in conducting materials under the influence of slowly varying magnetic fields \cite{Maxwell71,Jackson75}. 
When a material interface is present, the dynamics of these currents below the surface is associated with an evanescent field.
Mathematically, eddy currents are characterized by a purely imaginary frequency (overdamped oscillations) \cite{Intravaia09,Henkel10}. For metallic half-spaces overdamped modes manifest themselves in the (differential) mode density $\varrho^{T}(z_{a},\omega)$ as a branch-cut  (continuum of modes) in the complex frequency plane along the negative imaginary axis \cite{Intravaia10,Intravaia10a}. 
In the local (Drude) description, the branch-cut results from the square roots occurring in the surface impedances [Eqs. \eqref{SurfImpadance}] and it is bounded by the branch-points at $\omega=-\imath\Gamma$ and $\omega=-\imath\xi^{\rm lc}_{0}(p)$, where 
\begin{align}\label{xi0}
	\xi^{\rm lc}_0(p)
		\approx \Gamma
				\frac{p^2\lambdabar^{2}_{\rm p}}{1+p^2\lambdabar^{2}_{\rm p}}
\end{align} 
is the purely imaginary solution of $\omega^{2}\epsilon(\omega)/c^{2}-p^2=0$.
Both electromagnetic polarizations exhibit this feature. However, for the TM polarization the impact of the branch-cut is strongly reduced by the divergence of the dielectric function at zero frequency. Physically, this behavior is associated with the screening of surface charges arising from the orthogonal component of the electric field, which effectively decouples the dynamics of the field inside and outside the material. Consequently, the evanescent field generated in vacuum by the eddy currents is essentially TE polarized and therefore predominately magnetic.
In close connection, for the Drude model with $\Gamma(T)$ vanishing faster than $T$, calculations have shown that the inequality in Eq. \eqref{nonCommLimit} is related to a discontinuous behavior of $Z^{\rm TE}/Z^{\rm TE}_{0}$ in the complex plane, while $Z^{\rm TM}/Z^{\rm TM}_{0}$ is regular \cite{Intravaia08}.

In general, the contribution to the full mode density stemming from eddy modes can be isolated by utilizing a contour integration in the complex plane \cite{Intravaia10,Intravaia10a}
\begin{align}
	\eta^{T}(z_{a},\omega)
		&=-\int_{0}^{\Gamma}\frac{{\rm d}\xi}{\pi}~
			\frac{\xi}{\xi^2+\omega^2}
			\partial_{\xi}
			\mathrm{Im}[\Delta^{T}(z_{a},-\imath\xi+0^+)]\nonumber\\
		&=\partial_{\omega}\int_{0}^{\Gamma}
			\frac{{\rm d}\xi}{\pi}~
			\frac{\omega~\mathrm{Im}[\Delta^{T}(z_{a},-\imath\xi+0^+)]}
				 {\xi^2+\omega^2}.
\label{modeeddy}
\end{align}
By an expansion of the poles of the Green tensor related with Mittag-Leffler's theorem \cite{Markusevic88,Intravaia12b}, Eq. (\ref{modeeddy}) connects a Lorentzian spectrum centered at zero frequency to each overdamped mode $\omega=-\imath \xi$ along the branch cut \cite{Intravaia10}. As a result, we have that $\varrho^{T}(z_{a},\omega)=\eta^{T}(z_{a},\omega)+\varrho_{\rm other}^{T}(z_{a},\omega)$, where $\varrho_{\rm other}^{T}(z_{a},\omega)$ is the (differential) mode density corresponding to the remaining resonances in the system (cavity modes, bulk and surface plasmon-polaritons \cite{Haakh10,Intravaia07}). 
Since the Green tensor selects values of $p\sim 1/z_{a}$, Eq. \eqref{xi0} sets the characteristic frequency scale for eddy modes ($z_a\gg \lambdabar_{\rm p}$)
\begin{equation}\label{Eq:nulc}
	\omega\approx \nu^{\rm lc}=\frac{D}{z_a^2}
\end{equation}
and, consequently, $0<\omega<\nu^{\rm lc}$ gives the frequency region where $\eta^{T}(z_{a},\omega)$ is substantially different from zero.

Similarly to the plane-plane configuration \cite{Intravaia10a}, we find for the Drude model that the full mode density describing the magnetic Casimir-Polder interaction and that stemming from the eddy currents alone coincide at very low frequency, i.e. $\varrho^{T}(z_{a},\omega)\approx \eta^{T}(z_{a},\omega)$ for $\omega\to 0$. Note that the latter approximation becomes more accurate, the better the resonance from overdamped eddy modes and the atomic resonance $\omega\sim\omega_a$ can be separated in the mode spectrum. This is particularly true for $\Gamma/T\to 0$ as $T\to 0$ (see Appendix \ref{appendix1} for details).
Importantly, for the Drude model, $\eta^{T}(z_{a},0)$ diverges for $\Gamma\to 0$, but the product 
\begin{align}
	N^{T}_{\rm lc}=\eta^{T}(z_{a},0)\nu^{\rm lc}\approx \frac{\Phi^{T}(0)}{z^{3}_{a}},
\end{align}
stays constant (it does not depend on the diffusion coefficient $D$ and therefore on $\Gamma$). The quantity $N_{\rm lc}^T$ can be associated with the number of overdamped modes effectively participating in the interaction and is not modified even for vanishing collision-induced dissipation. In the next section we show how this feature contributes to the appearance of a non-vanishing entropy in the limit $T\to 0$. 


\section{Residual Entropy and Eddy Currents}
\label{Defect}

The dominance of eddy modes at low frequencies is also reflected in the behavior of the corresponding entropy contribution, $\mathcal{S}_{\rm e}(T)$. Upon replacing the system's full mode spectrum in the free energy of Eq. \eqref{FreeEnergy} with the scale defined in Eq. \eqref{modeeddy}, we can define the contribution to the magnetic Casimir-Polder free energy, $\mathcal{F}_e$, arising from the overdamped modes \cite{Intravaia10}.
For a constant dissipation rate, the corresponding entropy evaluates to $\mathcal{S}_{\rm e}(T\to 0)\approx(\pi^{2}/3) (k_{\rm B}^{2}/\hbar)\eta^{0}(z_a,0)T$. It linearly vanishes for $T\to 0$ and matches the behavior of the entropy due to the full mode spectrum (for more details see Appendix \ref{Entropyanalysis} and \ref{appendix1}).

If we now consider the case where $\Gamma(T)/T\to 0$ for $T\to 0$, we find that, due to the behavior of $\eta^{T}(z_{a},\omega)$, the modes contributing to the free energy have an energy $\hbar\omega\lesssim E^{\rm lc}=\hbar\nu^{\rm lc}\ll k_{\rm B}T$. This means that, even for a vanishing $T$, the relevant modes are \emph{always thermally excited} \cite{Intravaia09,Intravaia10,Intravaia10a}. In a broader perspective, we have the unusual circumstance that for $T\to 0$ a subset of the total system is \emph{behaving classically}.
Since the free energy per mode can be written as $
k_{\rm B}T\ln 2\sinh\left[\hbar\omega/(2k_{\rm B}T)\right]\approx k_{\rm B}T\ln [\hbar\omega/(k_{\rm B}T)]
$,  after a partial integration, we can estimate the free energy connected to eddy modes to
\begin{align}
	\mathcal{F}_{\rm e}(T)
		&=k_{\rm B}T\int_{0}^{\infty}{\rm d}\omega 
			\ln 2\sinh\left[\frac{\hbar\omega}{2k_{\rm B}T}\right]
			\eta^{T}(z_{a},\omega) \nonumber\\
		&\sim -k_{\rm B}T\int_{0}^{\nu^{\rm lc}}  
			\frac{{\rm d}\omega}{\omega}\int_{0}^{\omega}\text{d}\omega'
			\eta^{T}(z_{a},\omega') \nonumber\\
		&\xrightarrow{T\to 0} - k_{\rm B}T N^{0}_{\rm lc},
\label{sumofparticles}
\end{align}
where we use that $\nu^{\rm lc}$ vanishes faster than $T$ for $T\to 0$ and that $\eta^{T}(z_{a},\omega)$ is different from zero for $0<\omega\lesssim \nu^{\rm lc}$.
Relation (\ref{sumofparticles}) not only shows that the interaction entropy is extensive, but also that at low temperature, under the condition $\Gamma(T)/T\to 0$, the free energy stemming from eddy currents resembles the energy of an ensemble of \emph{identical classical oscillators} with a total number $\sim N^{0}_{\rm lc}$ \cite{Hanggi08,Intravaia09,Ingold09}. The corresponding entropy is thus constant and of the order of $k_{\rm B}$ (entropy per ``particle'') times the number of modes $N^{0}_{\rm lc}$ effectively participating in the interaction. This expression is consistent with the value obtained in Eq. \eqref{entropydefect} for $\mathcal{S}_{0}$. 
In other words we can say that in the limit $\hbar\Gamma(T)<k_{\rm B}T$, the eddy modes decouple from the external field (not even radiation damping occurs) and their finite-temperature disorder is partially frozen down onto the system's zero-temperature state. This state, which is reminiscent of some features of glassy systems but that curiously occurs in the limit of a vanishing damping, was dubbed \emph{Foucault glass} in previous work \cite{Intravaia09,Intravaia10a}.

Using Eq. \eqref{modeeddy}, a more careful calculation yields for $T\to0$ (see also Appendix \ref{appendix1})
\begin{align}
\label{FreeEnergyEddy}
	\mathcal{F}_{\rm e}(T)
		&\approx 
		k_{\rm B}T\int_{0}^{1}{\rm d}x~
		\frac{\mathrm{Im}[\Delta^{T}(z_{a},-\imath x \Gamma(T)+0^+)]}{2 x}.
\end{align}
For the Drude model in the limit $T\to0$, the integrand of the previous expression does no longer depend on the temperature and the corresponding entropy stays finite at $T=0$. Similarly to the procedure followed in Refs. \cite{Intravaia10,Intravaia10a}, using the analytic properties of the function $\Delta^{T}(z_{a},\omega)$, upon employing a shift of the integration contour toward infinity, we can show that $\mathcal{S}_{\rm e}(T= 0)=\mathcal{S}_{0}$. Indeed, the contour encircling the branch-cut characterizing the eddy currents can also be re-interpreted as a contour in the opposite direction around the remaining part of complex plane. We can then use the residue theorem and sum over the poles of the integrand. These are located at $x=0$ and at the poles of $\Delta^{T}(z_{a},\omega)$ for $\Gamma(T)\to 0$, which correspond to all the other modes of the system (excluding the eddy currents) in \emph{the non-dissipative limit of the material} (i.e. the plasma model). For the latter, the resulting expression is the same that one would have obtained in the high temperature (large distance) limit and we can write
\begin{align}
\label{FreeEnergyEddyContour}
	\mathcal{F}_{\rm e}(T)
		&\approx k_{\rm B}T\frac{\pi}{2} \left[
				  \Delta_{\rm plasma}^{0}(z_{a},0)
				  -\Delta^{T}(z_{a},0)\right].
\end{align}
The second term is arising from the pole in $x=0$. The subscript ``plasma'' in first term indicates that the limit $T\to 0$ in $\Delta^{T}(z_{a},\omega)$ is taken first and the limit $\omega\to 0$ afterwards.
Starting from Eq. \eqref{FreeEnergyEddyContour} we then arrive at an expression for the entropy, which is equivalent to that of Eq. \eqref{lowTemp}, showing that the eddy currents are entirely responsible for  the residual entropy evaluated for the full mode spectrum (see Fig. \ref{fig:S}, black dashed).


\section{Overdamped modes in the nonlocal description}
\label{NonlocalEddyModes}

For spatially nonlocal material models the above analysis must be revisited. In contrast to the Drude limit, at low frequency the Landau damping serves as an additional damping mechanism which will modify the eddy modes' properties and characteristic energy. 
Since it plays a central role in the behavior of the Casimir entropy at zero temperature, we focus in the following on the transverse electric impedance in Eq. \eqref{surfaceimpedanceS}.  
Substituting $x=p/k$, we can rewrite the surface impedance as
\begin{align}
	\frac{Z^{\rm TE}(\omega,p)}{Z_{0}^{\rm TE}(\omega,p)}
		=-\frac{2}{\pi}\int_{0}^{1}{\rm d}x~\frac{1}{\sqrt{1-x^2}}
		\frac{p\sqrt{p^2-\frac{\omega^2}{c^2}}}
		   	 {\epsilon_t\left(\omega,\frac{p}{x}\right)
		\frac{\omega^2x^2}{c^2}-p^2}.
\label{SI}
\end{align}
While $\epsilon_t\left(\omega,p/x\right)$ is a smooth function of $x$ over the integration range, $1/\sqrt{1-x^2}$ diverges for $x\rightarrow 1$. We can hence expand the dielectric function around $x=1$, obtaining at the leading order

\begin{figure}
  \centering
    \includegraphics[width=0.45\textwidth]{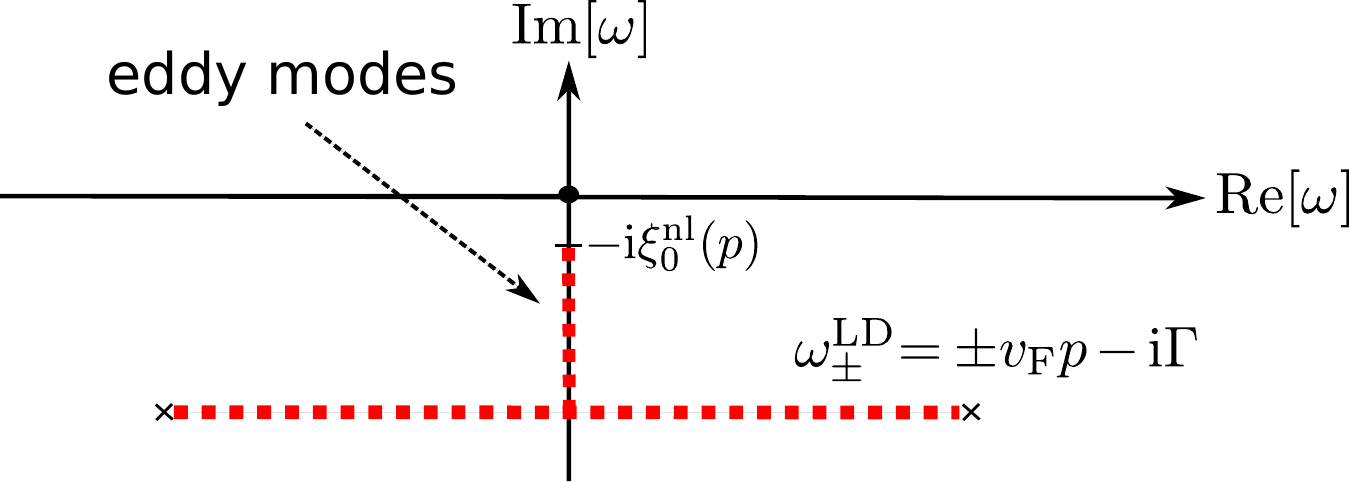}
    \caption{(Color online) Schematic visualization of the mode 
    density's branch-cut structure [see Eq. \eqref{SI2}] 
    for fixed wave vector in the complex frequency 
    plane using the spatially nonlocal model. 
    Eddy modes in the material are reflected by a branch-cut on the negative 
    imaginary axis constrained 
    by $\omega\in[-\imath\Gamma,-\imath\xi_0^{\rm nl}(p)]
    $. Parallel to the real frequency axis we find an additional branch-cut bounded 
    by $\omega_{\pm}^{\rm L}=\pm v_{\rm F}p-\imath\Gamma$ that is connected to Landau damping in 
    the material (see main text). Both cuts meet in $\omega=-\imath\Gamma$ and form 
    an upside-down T-shaped structure. \label{fig:complexF}}
\end{figure}
\begin{figure}
 \includegraphics[width=0.45\textwidth]{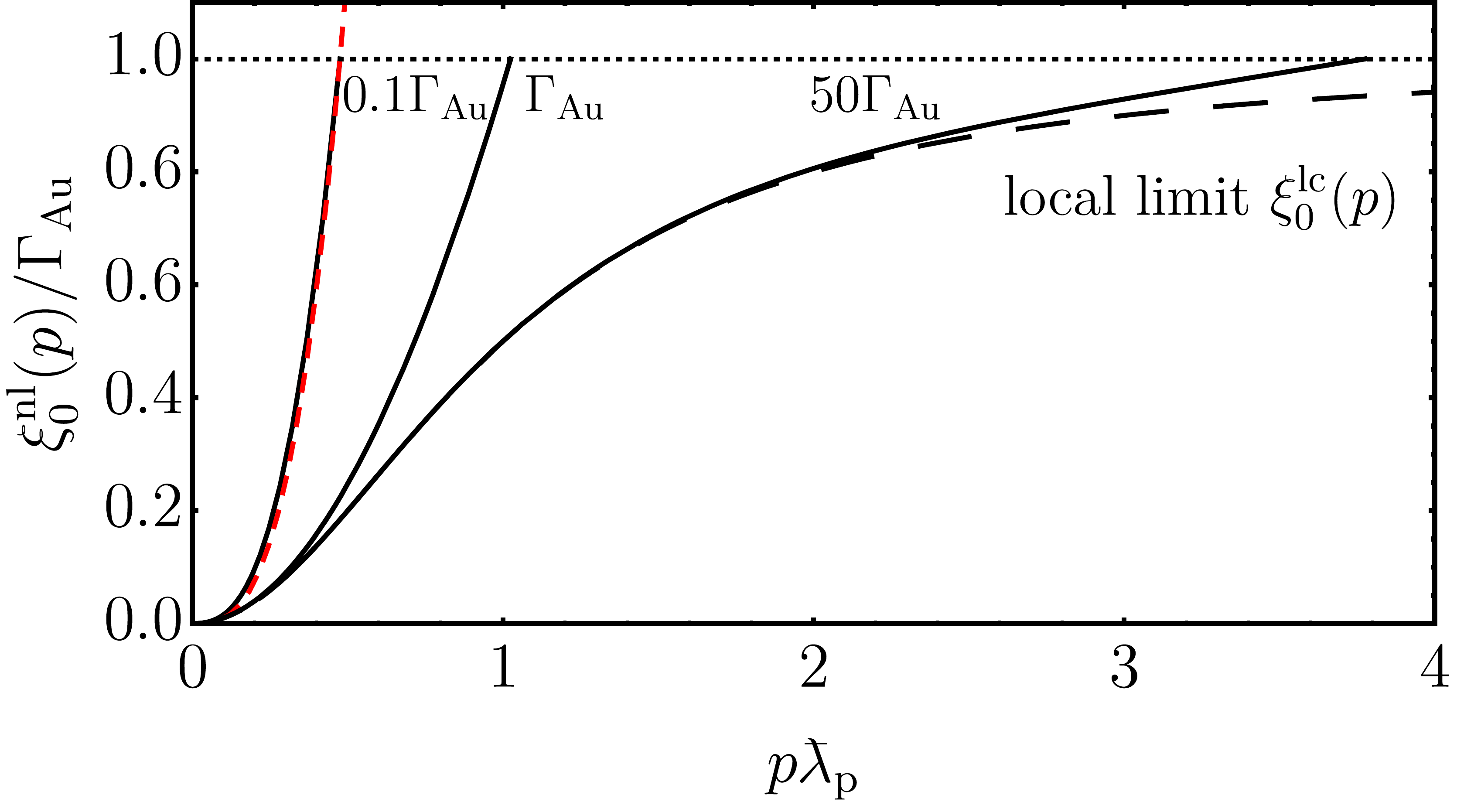}
  \caption{(Color online) We report $\xi_{0}^{\rm nl}$ as a function of $p
  \lambdabar_{\rm p}$ [see the parametric expression in 
  Eq. (\ref{parametricXi})] for 
  different values of the collision-induced dissipation $\Gamma$ (black, drawn 
  through). Parameters are chosen as in Fig. \ref{FreeEn} and we utilize the 
  dissipation rate of gold $\Gamma_{\rm Au}\sim 30$meV as a typical scale. For 
  positive wavevectors, $\xi_0^{\rm nl}(p)$ is strictly 
  bounded by the domain $0\leq
  \xi_0^{\rm nl}(p)\leq\Gamma$ (horizontal dotted line). 
  The larger the collision-induced dissipation rate $\Gamma$, the more 
  resembles $0\leq\xi_0^{\rm nl}(p)\leq
  \Gamma$ the local limit $\xi_0^{\rm lc}$ of Eq. (\ref{xi0}) (black, dashed). For 
  small dissipation rates the behavior of $\xi_0^{\rm nl}$ is well approximated by 
  Eq. (\ref{xinl}) (red, dashed).}
  \label{fig:Xi0}
\end{figure}
\begin{figure}
  \centering
    \includegraphics[width=0.43\textwidth]{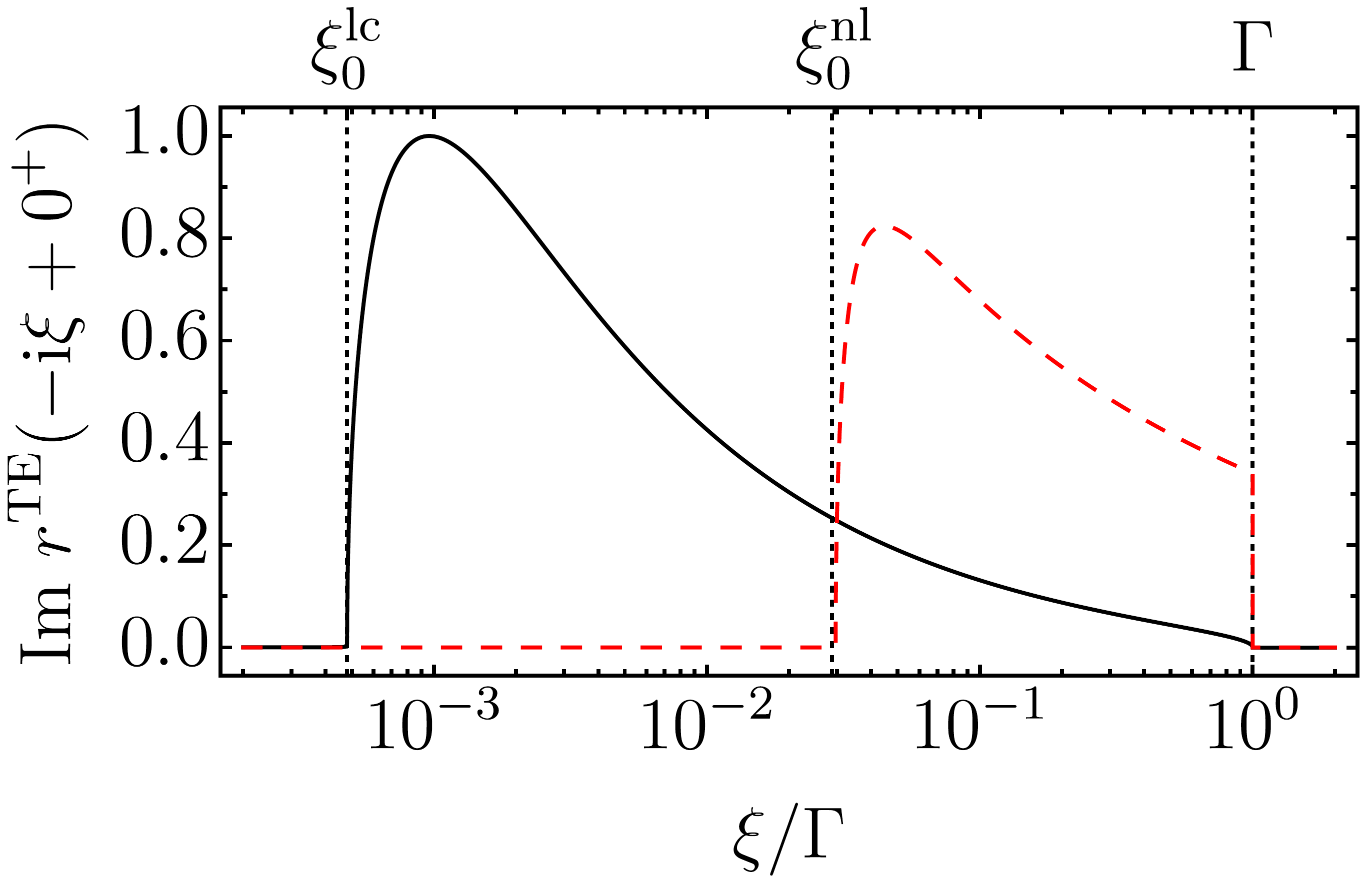}
    \caption{(Color online) Imaginary part of the transverse electric reflection 
    coefficient evaluated at the r.h.s. of its branch-cut on the negative imaginary 
    axis. The complex frequencies are measured in multiples of the dissipation rate 
    $\Gamma$. We depict the result for a local Drude model (solid line) and the 
    nonlocal SCIB model (dashed) both vanishing at their lower bound $\xi=\Gamma$. 
    The parameters are the same as in Fig. \ref{FreeEn} 
    and we fixed $\Gamma=10^{-5}$ eV as well as $p=1/\mu$m 
    in order to resolve the transition regime from local 
    to nonlocal response ($\lambdabar_{\rm e}^{-1}\approx 3/\mu$m). For fixed wave vector the branch-cut's upper bound $
    \xi_0^{\rm lc}$ of the local model is uniquely defined via the value of $\Gamma
    $. In the nonlocal case, however, the upper bound is given by $\xi_0^{\rm nl}$ 
    [see Eq. \eqref{xinl}].}\label{ImRef}
\end{figure}


\begin{align}
\label{SI2}
	\frac{Z^{\rm TE}(\omega,p)}{Z_{0}^{\rm TE}(\omega,p)}
		&\approx\frac{\sqrt{\frac{\omega^2}{c^{2}}-p^{2}}}
					 {\sqrt{\epsilon_t\left(\omega,p\right)
					 	\frac{\omega^2}{c^2}-p^{2}}
					 }.
\end{align}
Equation \eqref{SI2} reduces to Eq. \eqref{surfaceimpedanceS}  when $v_{\rm F}\ll\omega/p$ and/or $p\ell\ll 1$. It again shows that, since the electron's mean-free path diverges for a vanishing collision damping, using the Drude model in the limit $\Gamma(T)\to 0$ is equivalent to using a local description in a region where the SCIB in contrast predicts a substantial nonlocal response. 
Due to the form of $\epsilon_t\left(\omega,p\right)$, Eq. \eqref{SI2} exhibits a logarithmic branch-cut between the points $\omega_{\pm}^{\rm L}=\pm v_{\rm F}p-\imath \Gamma$.  As in the local case, Eq. \eqref{SI2} also features an algebraic branch-cut along the negative imaginary frequency-axis. The cut connects one branch-point still at $\omega=-\imath\Gamma$ with the other now located at $\omega=-\imath \xi^{\rm nl}_{0}(p)$. $\xi^{\rm nl}_{0}(p)$  is approximately given by the imaginary zero of $\epsilon_t\left(\omega,p\right)\omega^2/c^2-p^{2}$. Together, these branch cuts form an upside-down T-shaped structure (see Fig. \ref{fig:complexF}).
Physically, the appearance of the first cut parallel to the real-frequency axis is a direct consequence of the Landau damping in the semi-infinite bulk and derives from a singularity in the dynamics of the electronic fluid, which occurs when the quasi-particles have a speed $v\sim \omega/p$. Indeed, at zero temperature, the Fermi-Dirac distribution prescribes $\abs{v}\le v_{\rm F}$, which indicates a damping for $\omega\lesssim \pm v_{\rm F}p$ \cite{Manfredi05}. 
In our case an additional damping occurs due to a particle's finite collision rate $\Gamma$ and makes up for the second branch-cut.
It describes the generalization of the eddy current modes to the nonlocal case. 
For good conductors ($\ell \gg \lambdabar_{\rm p}$ and $\Gamma/\omega_{\rm p}\ll 1$), we obtain (see Appendix \ref{Branchpoint})
\begin{equation}
	\xi^{\rm nl}_{0}(p)\approx\Gamma^{\rm L}_{t}(p) p^{2}
		\lambdabar_{\rm p}^{2}
	,\quad  p\le \frac{1}{\lambdabar_{\rm e}(\ell)}\equiv\frac{1}{\ell}
		\left[\frac{3\pi}{4}\frac{\ell^{2}}{\lambdabar_{\rm p}^{2}}
		\right]^{\frac{1}{3}}.
\label{xinl}
\end{equation}
This is formally similar to Eq. \eqref{xi0}, but replaces $\Gamma \to \Gamma^{\rm L}_{t}(p)=4 v_{\rm F}p/(3\pi)$, showing that the Landau damping drives the diffusive process that characterize the dynamics of the eddy modes. 
The most important feature of the previous expression is, however, that the field can only diffuse for wavelengths $\lambdabar>\lambdabar_{\rm e}(\ell)$ due to spatial dispersion. As soon as the motion of the electrons becomes more and more ballistic and the mean free path $\ell$ becomes large, the eddy modes are no longer supported by the system -- they are effectively ``frozen out'' (see Fig. \ref{fig:Xi0}). 
This critical wavelength $\lambdabar_{\rm e}$ is equivalent to the condition
 $\Gamma=\xi^{\rm nl}_{0}(p)$, for which the branch-cut's upper and lower bounds coincide, leading to the disappearance of the corresponding eddy currents' contributions. 
In analogy to the local case we can define the typical excitation energy of eddy currents for spatially nonlocal conductors described by the SCIB model. Once again, since $p\sim 1/z_{a}$, the value $\xi^{\rm nl}_{0}$ corresponds to an energy scale 
\begin{align}\label{lambdae}
	E^{\rm nl}
		\approx \frac{4}{3\pi}
				\frac{\hbar v_{\rm F}
				\lambdabar_{\rm p}^{2}}{z_{a}^{3}}=\hbar\nu^{\rm nl}.
\end{align}
$E^{\rm nl}$ is directly related to the frequency scale encountered in the low-temperature behavior of the entropy within the nonlocal description [see Eq. \eqref{nonlocalfrequencyscale}].
From the point of view of the Casimir entropy the most important aspect is, however, that $\xi^{\rm nl}_{0}(p)$ does depend on the collision rate  only via the cutoff wavelength $\lambdabar_{\rm e}$.
For comparison, in the local Drude model, both branch-points are linearly proportional to $\Gamma$ and all eddy modes show the same dependence. 
Upon rescaling with respect to $\Gamma$, the branch-cut on the negative imaginary axis exists for all $0<p<\infty$. In the limit $\Gamma\to 0$ this effectively corresponds to infinitely many permanent current modes flowing through the plasma. Indeed, in the limit $\Gamma\to 0$ the mode density of all eddy modes mathematically collapses as $\lim_{\Gamma\to0}\eta^T(z_a,0)=\infty$, giving rise to the discontinuous behavior in Eq. \eqref{nonCommLimit}.
In the nonlocal case, it remains true that for $\Gamma(T)/T\rightarrow 0$ ($T\to 0$) the eddy currents are always excited. However, the Fermi-Dirac statistics (through the Landau damping) leads to a \emph{suppression} of the modes, which is different from the collapse in $\omega\to 0$ described above: The branch cut asymptotically disappears but no singular behavior occurs in $\omega\sim 0$. 
This is visible in Figs. \ref{fig:Xi0} and \ref{ImRef}, where we report the impact of the limit $\Gamma\to 0$ on the branch-point $\xi^{\rm nl}_{0}(p)$ and $r^{\rm TE}$. We see that, while in the local limit $\xi ^{\rm lc}(p)/\Gamma$ does not depend on the material damping, in the nonlocal case the range of wavevectors involved in the diffusive dynamics of the eddy currents shrinks with $\Gamma$. This eventually leads to the disappearance of the reflection coefficient's branch cut for $p\lambdabar_{\rm e}>1$ (see Fig. \ref{ImRef}).
As a consequence, for the magnetic Casimir-Polder free energy in the nonlocal case, the integrand of Eq. \eqref{FreeEnergyEddy} still depends on temperature and, since $\mathrm{Im}[r^{\rm TE}(-\imath x \Gamma(T)+0^+)]$ vanishes for $T\to 0$, $\mathcal{F}_{\rm e}(T)$ approaches zero faster than linearly ensuring $\mathcal{S}_{\rm e}(T\to 0)=0$. Repeating the analysis of Eq. \eqref{sumofparticles} one can still approximately write that $\mathcal{F}_e\sim-k_{\rm B}T N_{\rm nl}^T$ with $N_{\rm nl}^T=\eta^T(z_a,0)\nu^{\rm nl}$. The free energy is described again as an ensemble of identical classical particles, resembling the local description. However, in the nonlocal description, when the temperature decreases, the number of particles is reduced and $N_{\rm nl}^T\to 0$ for $T\to 0$.

\section{Thermodynamical properties of the overdamped modes}
\label{SingleMode}

Notwithstanding the oversimplifications inherent in the Drude model, in light of the previous results and of the numerous systems featuring a residual entropy at zero temperature \cite{Shirai18}, it is interesting to further investigate the origin of a finite value for $\mathcal{S}_{0}$. From the viewpoint of statistical physics, a vanishing entropy in the limit $T\to 0$ occurs only if the system has -- in this limit -- one and only one micro-state available, which is quantum-mechanically connected with the existence of a unique (non-degenerate) ground state \cite{Morse62,Kittel80}. If the ground state is degenerate, the system does not have a well-prescribed order at $T=0$ and a residual constant entropy appears \cite{Morse62}. As mentioned above, in condensed matter physics, such a behavior is typical of systems, where some amount of disorder remains at zero temperature or, equivalently, when there is lack of information about the state of the system at T=0 \cite{Kittel80,Shirai18}. 
The Casimir entropy  $\mathcal{S}_{0}\not=0$ points therefore to a situation where, for either the interacting or the non-interacting state or -- more likely -- for both of them, some form of disorder is still trapped in the system for $T\to 0$.

To understand how this aspect affects our system, it is convenient to analyze first the quantum-thermodynamical properties of a single overdamped mode of frequency $\xi$. As for the eddy currents, we assume that the mode frequency can depend on the temperature, i.e. $\xi\equiv\xi(T)$. For such a system the free energy can be written as  $\mathcal{F}_{\xi}=\mathcal{E}^{0}_{\xi}+\Delta \mathcal{F}_{\xi}$, where 
\begin{equation}
\label{eq:E0}
	\mathcal{E}^{0}_{\xi}
		=\frac{1}{\pi}\int_{0}^{\Lambda\to \infty}{\rm d}\omega\ 
		 \frac{\hbar\omega}{2}
		 \frac{\xi}{\omega^{2}+\xi^{2}}\sim-\hbar\xi\ln
		 \left[\frac{\xi}{\Lambda}\right]>0
\end{equation}
represents the ground state energy. In analogy to the free energy of an (over)damped harmonic oscillator \cite{Nagaev02,Hanke95,Hanggi06,Hanggi08,Ingold09}, this expression diverges and requires a regularization through a cut-off frequency $\Lambda$ \cite{Intravaia09,Note4}. 
The thermal contribution is given instead by
\begin{equation}
\label{eq:FXi}
	\Delta\mathcal{F}_{\xi}
		=\frac{k_{\rm B}T}{\pi}\int_{0}^{\infty}{\rm d}\omega\ 
		 \ln\left[1-e^{-\frac{\hbar\omega}{k_{\rm B}T}}\right]
		 \frac{\xi}{\omega^{2}+\xi^{2}}.
\end{equation}
The behavior of Eq. (\ref{eq:FXi}) in the limit $T\to 0$ strongly depends on the ratio $\xi/T$. We have that
\begin{equation}
	\Delta\mathcal{F}_{\xi}\xrightarrow{T\to 0}
		\begin{cases}
			-\frac{\pi k_{\rm B}^{2}T^{2}}{6\hbar\xi}<0 	& \xi/T\to\infty\\
			\frac{k_{\rm B}T}{2}
				\ln[\frac{\hbar\xi}{k_{\rm B}T}] <0 		&\xi/T\to 0
		\end{cases}.
\label{Limitoverdamped}
\end{equation}
The first case ($\xi/T\to\infty$) occurs, for instance, when the dispersive mode frequency $\xi$ approaches a nonzero constant for $T\to 0$. The second is typical for $\xi\propto \Gamma(T)$ with $\Gamma(T)\propto T^{m}$ and $m>2$ such as in the perfect crystal limit. Let us consider next the corresponding partition function $Z(\beta)=\exp[-\beta \mathcal{F}_{\xi}]$ and the density of states $\rho_{\rm dos}(E)$ defined by 
\begin{equation}
	Z(\beta)=\int_{0}^{\infty}{\rm d}E\,
			 \rho_{\rm dos}(E) e^{-\beta E},
\end{equation}
where $\beta=(k_{\rm B}T)^{-1}$
\cite{Note5}.
In the case of a temperature-independent $\xi$  ($\xi/T\to\infty$ for $T\to 0$), the density of states reads,
\begin{equation}
	\rho_{\rm dos}(E)
		=\delta(E-\mathcal{E}_{\xi}^{0})
		+\sigma(E-\mathcal{E}_{\xi}^{0})
		\theta(E-\mathcal{E}_{\xi}^{0}),
\end{equation}
where $\theta(x)$ is the Heaviside function and $\sigma(x)$ is a function such that $\sigma(x)\to \pi/(6\hbar\xi)$ for $x\to 0$. The density of states shows that the system has a unique ground state at energy $E=\mathcal{E}_{\xi}^{0}$ plus a continuous spectrum of excited modes starting at $E=\mathcal{E}_{\xi}^{0}$ \cite{Hanke95}. Accordingly, the corresponding entropy vanishes at $T=0$.

However, if $\xi/T\to 0$ the above analysis yields drastically different results. Indeed, both $\mathcal{E}_{\xi}^{0}$ and $\Delta\mathcal{F}_{\xi}$ depend on temperature and the difference between ground state energy and thermal correction looses its meaning. In addition, in the limit $T\to 0$ in Eq. \eqref{Limitoverdamped} one has $\Delta\mathcal{F}_{\xi}\gg \mathcal{E}_{\xi}^{0}$ so that the partition function takes the form
\begin{equation}
	Z(\beta)\xrightarrow[\hbar\xi\beta\to 0]{\beta\to \infty}
	\left[\frac{1}{\hbar\xi\beta}\right]^{\frac{1}{2}}.
\label{Zeddy}
\end{equation}
Equation (\ref{Zeddy}) exhibits certain similarities with the partition function of a free classical oscillator or a particle trapped in a one-dimensional box \cite{Hanggi06,Hanggi08,Ingold09}.
Notice, however, that the expression in Eq. \eqref{Zeddy} diverges in the limit $T\to 0$ while a for a free particle it goes to zero. This can be seen as a direct consequence of the fact that the ``size'' of the oscillator $L\propto \sqrt{\hbar/ \xi}$ becomes much larger than the thermal de Broglie wave length $\lambda_{\rm th}\propto \sqrt{\hbar^{2}\beta}$  \cite{Huang03}, i.e.
\begin{equation}
Z(\beta)\approx \frac{L}{\lambda_{\rm th}} \to \infty.
\end{equation}
This is the range where quantum effects must become important, while the description above shows that the diffusive currents are (one could say incoherently) behaving as a classical system.

Let us now specify the calculation to the overdamped modes occurring in our system when $\Gamma(T)/T\to 0$ and the Drude model is used. Due to the properties of the branch-points discussed in Sec. \ref{LocalEddyModes} all the mode frequencies are proportional to the collision rate. As in Eq. \eqref{FreeEnergyEddy}, we can write $\xi=x \Gamma(T)$ and multiply the partition function in Eq. \eqref{Zeddy} by $\left(\hbar\Gamma\beta\right)^{\frac{1}{2}}$. This amounts to performing a common shift in the energy of all modes, which, due to its differential nature, does not affect the thermodynamical properties of the corresponding Casimir interaction. 
The relevant partition function of a single overdamped mode, consequently, is given by
\begin{equation}\label{rhodos}
	\tilde{Z}(\beta)
	\xrightarrow[\hbar\xi\beta\to 0]{\beta\to \infty}
			\left[\frac{1}{x}\right]^{\frac{1}{2}}
	\Rightarrow \tilde{\rho}_{\rm dos}(E)
	\xrightarrow{E\to 0}\left[\frac{1}{x}\right]^{\frac{1}{2}}\delta(E).
\end{equation}
We find a density of states that describes a system with a ground state having an effective degeneracy $\left[1/x\right]^{\frac{1}{2}}$. 
Since we have not singled out a specific overdamped mode, Eq. (\ref{rhodos}) characterizes all of these modes and, in particular, they all have a common ground state. Only the degeneracy, which grows as the inverse square root of $\xi/\Gamma$, is quantitatively different from mode to mode. The contribution to the Casimir interaction stemming from these resonances is obtained by summing (integrating) over all these modes using the differential mode density $\mu(x)=-\lim_{T\to0}\partial_{x}\mathrm{Im}[\Delta^{T}(z_{a},-\imath x \Gamma(T)+0^+)]$. For the Drude model, since all modes are proportional to the collision rate, the parametrization discussed above leads to a value of $\mu(x)$ which is nonzero.
The resulting expression is the free energy given in Eq. \eqref{FreeEnergyEddy} and the finite entropy $\mathcal{S}_{0}$ can therefore be interpreted as resulting from the \emph{differential} degeneracy $\exp[-\mathcal{S}_{0}/k_{\rm B}]$ of the system's ground state.
The value is distance-dependent because the modal spectral density in the interacting configuration is different from that in the non-interacting one. As a result the differential number of modes behaving classically changes as a function of $z_{a}$.
Our findings are consistent with Eq. \eqref{sumofparticles}, where the free energy due to the eddy currents has be interpreted as that of an effective number $N^{0}_{\rm lc}$ of identical classical oscillators.

In the nonlocal case the frequency of the overdamped modes depends on the collision rate in a way that is different from the local case. It is still possible to write $\xi=x \Gamma(T)$, due to the cutoff in $p$ [see Eq. \eqref{xinl}], but the density $\mu(x)$ vanishes this time with $T\to 0$. 
While the eddy currents in the nonlocal case preserve their characteristics (e.g. thermally excited at low temperature, degenerate ground state), due to an intrinsic mechanism related to the SCIB description, their number $N_{\rm nl}^T$ is drastically reduced to zero for $T=0$ (see also the argument at the end of the previous Section and discussion below). 


\section{Magnetohydrodynamics and London's theory}
\label{MHD}

It is insightful to add yet another (hydrodynamic) perspective to our analysis. As we have mentioned several times in the foundational sections, a metal can essentially be described as a plasma of electrons. In addition, the eddy currents are related to a diffusive dynamics of the electromagnetic field within a dissipative metal (see for example Sec. \ref{SingleMode}). In classical plasma physics this phenomenon is known as \textit{magnetic viscosity} and can be regarded as a simple consequence of the magnetohydrodynamic equations of motion \cite{Bittencourt13}. It also shares interesting connections with one of the oldest phenomenological descriptions of superconductivity, i.e. the London theory \cite{London35,Tinkham04,Schmalian10}.

For a spatially local homogeneous non-magnetic metal we can write the equation of motion for the current $\mathbf{j}(\mathbf{r},t)$ in terms of the electric field $\mathbf{E}(\mathbf{r},t)$ as 
\begin{equation}
	(\partial_{t}+\Gamma)\mathbf{j}(\mathbf{r},t)
	=\frac{\epsilon_{0}c^{2}}{\lambdabar_{\rm p}^{2}}\mathbf{E}(\mathbf{r},t),
\label{DrudeCurrent}
\end{equation}
which is known for leading to the Drude permittivity. 
At low frequencies the partial derivative can be neglected and the current fulfills Ohm's law. 
It is important to note that Eq. \eqref{DrudeCurrent} assumes a local relation between the field and the current, which can be justified only if the field is essentially homogeneous. 
As we discussed earlier, some difficulties can appear due to eventual interfaces \cite{Tinkham04}.
Upon combining the above expression with the Maxwell equations and neglecting the displacement current, we obtain for the magnetic field $\mathbf{B}(\mathbf{r},t)$,
\begin{equation}
	\partial_t\mathbf{B}(\mathbf{r},t)
	=(\lambdabar_{\rm p}^{2}\partial_{t}+D)\nabla^{2}\mathbf{B}(\mathbf{r},t),
\label{MagneticFieldEq}
\end{equation}
where $D=\Gamma \lambdabar_{\rm p}^{2}$ is once again the diffusion coefficient  introduced in Sec. \ref{nonlocality} (in the framework of magnetohydrodynamics sometimes called magnetic viscosity coefficient \cite{Bittencourt13}). At low frequency (i.e. the Ohmic regime), we can neglect the time derivative on the r.h.s. of Eq. \eqref{MagneticFieldEq} relative to the diffusion coefficient. We then have $\partial_t\mathbf{B}(\mathbf{r},t)=D\nabla^{2}\mathbf{B}(\mathbf{r},t)$, showing that the magnetic field effectively obeys a \emph{diffusion equation}.
The corresponding expressions for the (eddy) currents as well as the electric field associated to this diffusive magnetic field can be derived from Maxwell equations and one can show that inside the material the fields are predominately magnetic in nature \cite{Jackson75,Henkel10,Note6}.  
In magnetohydrodynamics, the limit for $D\to 0$ (occurring for example when $\Gamma\to 0$) in the diffusion equation of the magnetic field $\mathbf{B}$ corresponds to a regime where the plasma is behaving as a highly conducting fluid. The magnetic field lines, instead of diffusing through the plasma, are \emph{frozen} within the material \cite{Bittencourt13}. 
The latter can be concluded from the following consideration. For a vanishing collision rate, Eq. \eqref{DrudeCurrent} is replaced by the equation $\partial_{t}\mathbf{j}(\mathbf{r},t)=(\epsilon_{0}c^{2}/\lambdabar_{\rm p}^{2}) \mathbf{E}(\mathbf{r},t)$ called the ``acceleration equation'' by F. and H. London \cite{London35,Tinkham04}. 
The acceleration equation was used to describe the existence of stationary currents in a superconductor \cite{London35,Tinkham04} and, when combined with Maxwell equations,  leads to the expression 
\begin{equation}
	\nabla^{2}[\mathbf{B}(\mathbf{r},t)-\mathbf{B}_{0}(\mathbf{r})]
	=\frac{1}{\lambdabar_{\rm p}^{2}}
	 [\mathbf{B}(\mathbf{r},t)-\mathbf{B}_{0}(\mathbf{r})].
\label{London}
\end{equation}
The main consequence of Eq. (\ref{London}) is that, as observed in a superconductor, any external field $\mathbf{B}(\mathbf{r},t)$ can penetrate the material only within a length of the order of $\lambdabar_{\rm p}$. More interesting for us is the field $\mathbf{B}_{0}(\mathbf{r})$, which in London's theory represents a frozen magnetic field: A relic describing the ``memory of the field existing in the metal'' \emph{before} the transition to the non-dissipative limit \cite{London35}. By implication, the total field deep inside the material can be nonzero if it was nonzero at some initial time.
Observing that the Meissner effect always leads to the expulsion of the magnetic flux from the superconductor (perfect diamagnetism), the London brothers concluded that the acceleration equation is not compatible with a description of superconductivity. Instead, they suggested a modification of the equations which is effectively equivalent to the assumption of the additional boundary condition $\mathbf{B}_{0}(\mathbf{r})=0$. 

\begin{figure}
 \includegraphics[width=0.45\textwidth]{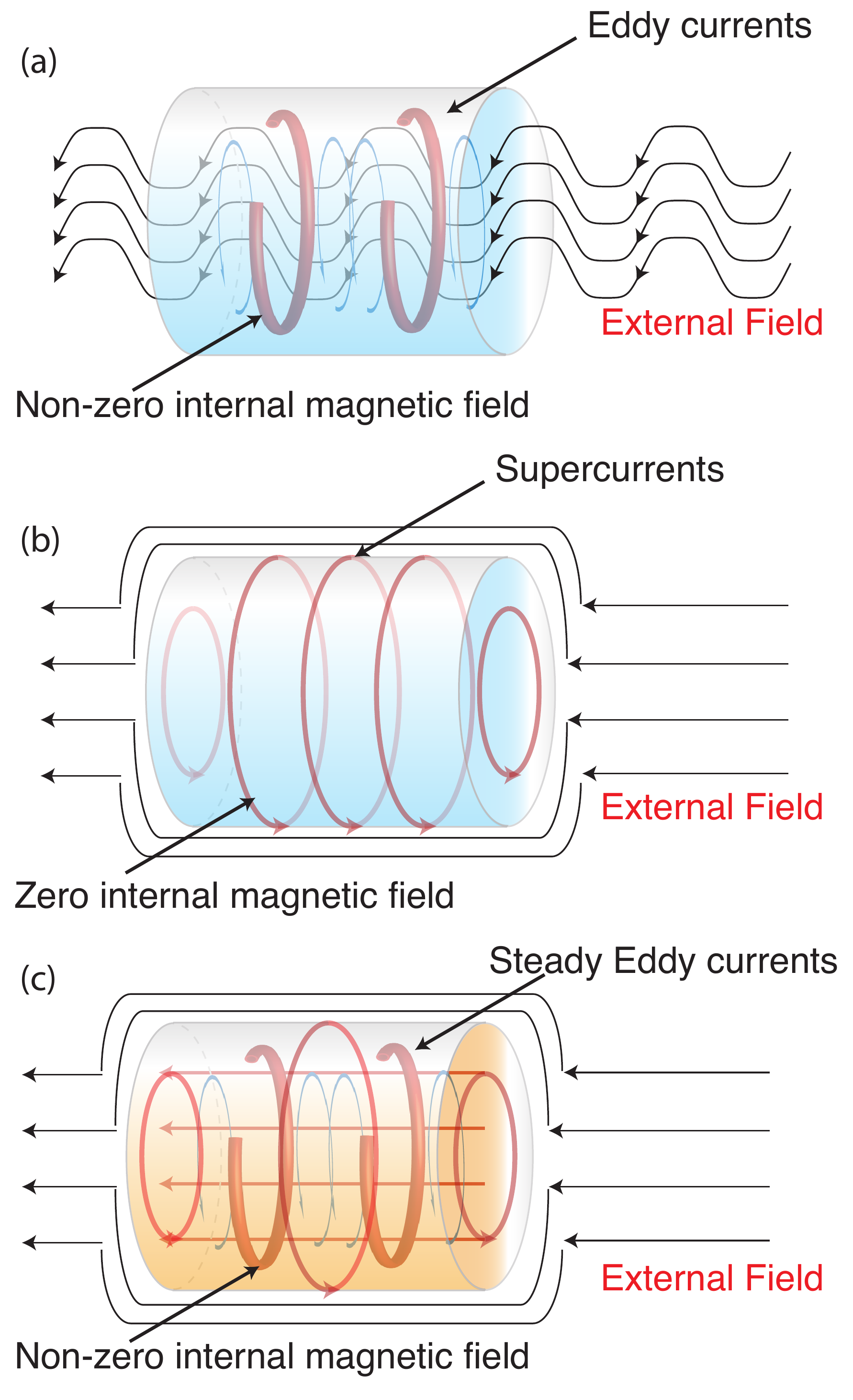}
  \caption{(Color online) Illustration of low-frequency magnetic field 
  characteristics associated with conducting materials.
  Panel a) A Drude metal is transparent for low-frequency 
  magnetic fields (Bohr--van 
  Leeuwen theorem \cite{Leeuwen21}). 
  Panel b) A superconductor will actively exclude 
  any magnetic field present when it undergoes the phase transition to the 
  superconducting state (Meissner effect). In agreement with the London theory, a 
  non-dissipative version of the Drude model (sometimes called plasma model) which 
  assumes a vanishing field inside the material is a simple description of a 
  superconductor \cite{London35}. 
  Panel c) If a (classical) conductor 
  already is penetrated by a magnetic field and is then cooled through the 
  transition to a zero-resistance state (ideal conductor), the magnetic field is 
  expected to maintain in place (solution discarded by the London theory) 
  \cite{Henyey82}. \label{Meissner&Co}}
\end{figure}

Let us now return to our original system. We have seen that using the Drude model for a dissipative (local) description of the material \emph{first} and then taking the limit $\Gamma\to 0$ \emph{afterwards} induces a screening which prevents any additional field from penetrating into it, effectively decoupling the internal material dynamics from the external field. Matter cannot dissipate energy even through radiation and a static memory of the initial magnetic field is preserved inside it (see Fig. \ref{Meissner&Co}). Imposing that the initial field is zero inside the material is essentially equivalent to adding a further constraint (not included in the Drude model description), which emulates the Meissner effect in superconductivity. In these terms, a Drude model, where the dissipation rate is removed from the very beginning (plasma model) and the internal field is set to zero, is nothing but a simple description of a superconductor \cite{Intravaia09,Bimonte10}.

Since we are dealing with fluctuating fields, there is one more subtlety to be discussed.
The relic field always has a zero mean value, $\langle\hat{\mathbf{B}}_{0}(\mathbf{r})\rangle=0$, but nonzero correlation $\langle\hat{\mathbf{B}}_{0}(\mathbf{r})\hat{\mathbf{B}}_{0}(\mathbf{r}')\rangle\not= 0$. In order to see more clearly how a non-vanishing correlation occurs, it is convenient to use the fluctuation-dissipation theorem for the currents inside the material
\begin{multline}
	\left\langle\hat{\mathbf{j}}(\mathbf{r},\omega)
				\hat{\mathbf{j}}(\mathbf{r}',\omega')
	\right\rangle_{\rm sym}=\\
	2\pi\hbar\omega^{2}\coth\left[\frac{\hbar\omega}{2k_{\rm B}T}\right] 
	\mathrm{Im}[\epsilon(\omega,\mathbf{r},\mathbf{r}')]\delta(\omega+\omega'),
\end{multline}
where we considered the symmetric correlation function. For a bulk described by the Drude model we have that
\begin{equation}
	\mathrm{Im}[\epsilon(\omega,\mathbf{r},\mathbf{r}')] 
	=\frac{\Gamma}{\omega}
	\frac{\omega_{\rm p}^2}{\omega^{2}+\Gamma^{2}}
	\delta(\mathbf{r}-\mathbf{r}').
\end{equation}
If we now consider the limit $T\to 0$ with $\Gamma(T)\to 0$, but $\Gamma(T)/T\to 0$, the current-current correlation tensor becomes time-independent and takes the form
\begin{equation}
	\langle\hat{\mathbf{j}}(\mathbf{r},t)
		   \hat{\mathbf{j}}(\mathbf{r}',t')
	\rangle_{\rm sym}
		\sim \frac{\omega_{\rm p}^2}{2\pi}k_{\rm B}T\delta(\mathbf{r}-\mathbf{r}').
\label{CurrentsCorr}
\end{equation}
Eq. (\ref{CurrentsCorr}) is almost identical to that obtained in the limit $T\to \infty$ (classical limit) and constant $\Gamma$, where, however, a factor $e^{-\Gamma\abs{t-t'}}$ appears. The two expressions are identical for $\Gamma \to 0$.
Within the Green tensor formalism one can also show that the field correlator is nonzero and time-independent in the zero-temperature limit.
In the limit $\Gamma\to 0$, Eqs. \eqref{MagneticFieldEq}-\eqref{CurrentsCorr} hence reveal that the static (zero-frequency) component of the field fluctuations, which were originally existing below the surface, are trapped inside the bulk and supported by permanent currents. These currents evolve from the eddy currents in the limit of vanishing dissipation and behave classically even at low temperature and despite the fact that no additional fields can penetrate the material.

The free energy in Eq. \eqref{FreeEnergyEddy} and the corresponding entropy can be regarded as thermodynamically characterizing the state of the relic field (what has been called Foucault glass above). In fact, Eq. \eqref{FreeEnergyEddy} can be seen as a mathematical consequence of the Bohr--van Leeuwen theorem \cite{Leeuwen21}. The theorem states that, in classical systems, a consistent application of statistical mechanics and classical mechanics does not allow for magnetic effects. Effectively, this means that matter and transverse field components decouple. The free energy of the total system becomes the sum of the free energy of the two subsystems material and field \cite{Alastuey00,Buenzli05,Buenzli06,Bimonte09}. 
As we saw, the limit $T\to 0$ with $\Gamma(T)/T\to 0$ corresponds to a regime where the currents in the metal are behaving classically even at low temperature.
In this unusual regime an application of the Bohr--van Leeuwen theorem gives that, due to the decoupling between matter and field, the magnetic contribution to free energy arising from the permanent currents within the material is quantitatively equivalent but opposite in sign to the magnetic part of classical free energy of the electromagnetic field. Note that the mode density of the latter must be obtained in the limit $\Gamma\to 0$ \cite{Intravaia10,Intravaia10a}, i.e. for a non-dissipative material description.  
Hence, in this limit the eddy current's free energy is formally equivalent to the (negative) Casimir free energy obtained for the plasma model in the high temperature limit (see i.e. Eq. \eqref{FreeEnergyEddy} and discussion at the end of Sec. \ref{Defect}). This last value must be corrected to exclude the contribution stemming from the electric field related to TM polarizations (the second term in the r.h.s. of Eq. \eqref{FreeEnergyEddyContour}).

Notice that the arguments outlined above do not apply to the permittivities given in Eqs. \eqref{permittivities}. It reveals indeed that the main difference between the spatially local and nonlocal material descriptions does not arise from the nonlocality itself but rather and more generally, from the statistical behavior of the system. 
Even though a nonlocal interaction is relevant for describing the behavior of the eddy currents and can lead to the disappearance of the entropy defect,  not all nonlocal descriptions are equivalent. 
For instance, hydrodynamic extensions of the Drude model (see for example \cite{Intravaia15a,Moeferdt18}) also lead to a finite entropy defect at zero temperature \cite{Dalvit08c,Decca09b,Dalvit09a}. It is using a full quantum-statistical description for the equilibrium distribution of the electrons in the plasma which drastically modifies the behavior of the eddy currents at low temperature and leads to their disappearance. 
In the specific case of Eqs. \eqref{permittivities}, even if $\Gamma(T)/T\to 0$ for a vanishing temperature, the current-current correlation cannot behave according to classical statistics and, due to the Fermi-Dirac distribution, must always display some degree of ``quantumness'' (preventing also the application of the Bohr--van Leeuwen theorem). 
Within the present description, we can therefore say that the mechanism leading to the vanishing entropy at $T=0$ is rather the quantum-statistical description of the electrons than the Landau damping, although the latter is always a consequence of the former.


\section{Discussion and conclusions}
\label{Conclusions}

In the present manuscript, we have isolated and studied the thermodynamic 
properties of overdamped resonances (eddy currents), with special focus on their 
role in atom-surface interactions. 
In particular, we have analyzed the behavior of the magnetic Casimir-Polder 
entropy in the limit of low temperatures for the spatially local (Drude) model 
and a nonlocal description of a metal based on the semi-classical Boltzmann-Mermin
description, the SCIB model. 
If the electrons' collision rate vanishes for $T\to 0$ (e.g. perfect crystal 
model) as a power law of temperature with an exponent larger than one, the previous 
material descriptions give results that qualitatively differ in the low-temperature
regime: While the \textit{nonlocal} description leads to a vanishing entropy, 
the \textit{local} model provides a nonzero distance-dependent entropy. 
We have shown that the behavior 
of the Drude model 
is associated with an unusual regime where the eddy currents decouple from the external radiation, become undamped and behave classically even at low temperatures. This means that they are characterized (i) by an excitation energy which is always lower than 
$k_{\rm B}T$ and (ii) by a degenerate ground state. For a local description 
in the limit $T\to 0$ and $\Gamma(T)/T \to 0$ all overdamped frequency modes 
collapse toward the ground state following classical (Maxwell-Boltzmann) statistics and inheriting the constraints of this description in terms of statistical mechanics. Since the (differential) spectral mode density depends on the atom-surface separation, the number of eddy currents effectively follows this behavior giving rise to the non-vanishing distance-dependent entropy.

Overdamped modes also occur within the nonlocal description. Nevertheless, 
the inclusion of spatial dispersion deeply modifies their structure and behavior. 
The most important change is the appearance of a cut-off 
in the dispersion relation in one of the two branch points defining the cut 
along the negative imaginary axis that represent the eddy currents in the 
complex-frequency plane. We have connected the appearance of the cut-off with 
the Landau damping occurring in the nonlocal system, but ultimately it is the 
underlying Fermi-Dirac statistics that characterizes the spatial dispersion 
and that leads to a vanishing entropy at zero temperature. Physically, this 
amounts to limitations in the diffusive dynamics of the currents when the 
electrons' ballistic regime becomes dominant. As a result, the eddy current mode 
density is very different from the local case and disappears in the limit 
$T\to 0$ and $\Gamma(T)/T\to0$ (instead of collapsing towards $\omega\sim 0$), leading to
a vanishing Casimir entropy for $T\to 0$.
Eddy currents are diffusive bulk currents and not surface currents \cite{Henkel10,Henkel19b}. They are therefore sensitive to the material's optical response in the region inside the object itself. Our analysis highlights that, in addition to the common focus on the surface's properties,  
attention must be also devoted to the
bulk's properties, since they can become relevant in determining the properties of the Casimir(-Polder) interaction.

We have related our results to the characteristics of the electromagnetic 
field within a plasma described by the standard magnetohydrodynamic
theory \cite{Bittencourt13}. For vanishing viscosity (dissipation), the magnetic 
flux is trapped within the material supported by currents which are 
the non-dissipative static ($\omega\sim 0$) version of the eddy currents. 
While any additional magnetic field can penetrate the material only to within 
a distance comparable to the reduced plasma wavelength, a relic field disorder 
can still pervade the material and preserves a certain memory of the diffusive 
dynamics occurring for finite dissipation. It is the free energy and the 
entropy of this relic field which is responsible for the finite Casimir entropy at 
$T=0$. Such a situation can only occur if the currents 
within the plasma behave classically even at low temperatures and their density
is high enough. 
It is prohibited, however, once certain statistical properties (in our case the Pauli 
exclusion principle via the Fermi-Dirac statistics) enter the description of 
the system. 

Finally, we would like to note that the above description can also account 
for certain properties of the Casimir free energy in metallic films 
\cite{Klimchitskaya15c}. 
Here, it was shown that in the limit of a diverging plasma frequency, 
for which the perfect electric conductor limit should be recovered, the 
Drude model and its non-dissipative version, the plasma model, behave quite 
differently. At high temperatures, the plasma model gives rise to a vanishing 
free energy while it stays constant for the Drude model. Once again, a diverging 
plasma frequency implies a transition of the material from a diffusive to a highly 
conducting state ($D\to 0$ for $\lambdabar_{\rm p}\to 0$) and the constant free 
energy corresponds to that of the relic field disorder trapped in the material. 
As discussed in the main text, a description in terms of the plasma model is 
consistent with the description of a metal in the superconducting state for 
which the field inside the material has been expelled because of the Meissner 
effect. The different results for the Casimir free energy of metallic films 
can, therefore, be explained from a lack of perfect diamagnetism for the perfect 
electric conductor limit obtained from the Drude model.


\section{Acknowledgments}

We thank C. Henkel and D.A.R. Dalvit for very useful discussions on the topic of this paper.
We acknowledge support by the LANL LDRD program and by the Deutsche 
Forschungsgemeinschaft (DFG) through project B10 within the Collaborative 
Research Center (CRC) 951 Hybrid Inorganic/Organic Systems for Opto-Electronics 
(HIOS). 
F.I. further acknowledges financial support from the DFG through the DIP program (Grants
FO 703/2-1 and SCHM 1049/7-1).
D.R. would like to thank the CNLS for the hospitality and acknowledges financial support
by the CNLS and the German Academic Exchange Service DAAD (PROMOS program). 
KB would like to thank Peter W\"olfle and Ralph von Baltz for numerous enlightening discussions. 


\appendix

\section{The function $\Delta^T(z_a,\omega)$}
\label{appendixentropylow}

This appendix is devoted to the analysis of the function $\Delta^T(z_a,\omega)$ obtained for the magnetic Casimir-Polder interaction, when we use the nonlocal description of the metal. We will focus on the non-retarded region $z_{a}\ll \lambda_{a}$. From Eqs. \eqref{RIGreen} and \eqref{phi} we can write
\begin{equation}
	\Delta^T\left(z_a,\omega\right)
		\approx -\frac{\Phi^{T}(\omega)}{z_{a}^{3}}
			    \int_{0}^{\infty}{\rm d}y~y^2e^{-y}r^{\rm TE}
			    \left(\omega,\frac{y}{2z_{a}}\right),
\label{start}
\end{equation}
where we make the change of variable $y=2z_{a}p$. For the local description of the metal we can use Eq. \eqref{apprTE} and obtain
\begin{equation}
	\Delta^T\left(z_a,\omega\right)
		\approx -\frac{\Phi^{T}(\omega)}{z_{a}} 
				\frac{\omega^{2}}{c^{2}}[\epsilon(\omega)-1].
\end{equation}
In the nonlocal case, using Eqs. \eqref{reflectioncoefficients} and \eqref{surfaceimpedanceS}, the reflection coefficient can be written as
\begin{align}
	r^{\rm TE}(\omega,p)
		&\approx \frac{1}{2}
				 \left[\frac{Z^{\rm TE}(\omega,p)}
				 			{Z_{0}^{\rm TE}(\omega,p)}
				 	   -1\right]\nonumber\\
		&\approx \frac{1}{2}
				 \left[\int_{0}^{1}{\rm d}x~
				 	   \frac{\frac{2}{\pi}}{\sqrt{1-x^2}}
					   \frac{1}{1-\epsilon_t
					   \left(\omega,\frac{p}{x}\right)
					   \frac{\omega^2x^2}{p^2 c^2}}-1
				 \right]\nonumber\\
		&= \frac{1}{2}\int_{0}^{1}{\rm d}x~
		   \frac{\frac{2}{\pi}}{\sqrt{1-x^2}}
		   \frac{\epsilon_t\left(\omega,\frac{p}{x}\right)
		   		 \frac{\omega^2x^2}{p^2 c^2}}
		   	    {1-\epsilon_t\left(\omega,\frac{p}{x}\right)
		   \frac{\omega^2x^2}{p^2 c^2}},
\label{refTEApp}
\end{align}
where we have used that
\begin{equation}
	\int_{0}^{1}{\rm d}x~\frac{\frac{2}{\pi}}{\sqrt{1-x^2}}=1.
\end{equation}
From Eqs. \eqref{permittivities}, in the limit $|\omega|\ll \Gamma$, we can use the following approximation for the dielectric function
\begin{align}
\epsilon_t\left(\omega,\frac{y}{2z_{a}x}\right)
		&\approx\imath\frac{\omega_{\rm p}^2}{\omega\Gamma}
		        g_t\left(\imath\frac{z_{a}}{\ell}\frac{2x}{y}\right),
\end{align}
where $\ell=v_{\rm F}/\Gamma$ is again the mean-free path.
Since the largest contributions arise from $x,y\sim 1$, the ratio $z_{a}/\ell$ is determining the limiting behavior of the function $g_{t}(u)$. As explained in the main text for $z_{a}\gg \ell$, one has $g_{t}\sim 1$  and we recover the local result in the corresponding limit.
In the opposite case ($|u|\to 0$) we have
\begin{align}
\epsilon_t\left(\omega,\frac{y}{2z_{a}x}\right)
		&\approx \imath\frac{\omega_{\rm p}^2}{\omega}
				 \frac{3\pi}{2}\frac{ z_{a}}{v_{\rm F}}\frac{x}{y}.
\label{epsapprox}
\end{align}
Notice that the previous expression is valid only if
\begin{equation}
	\frac{z_{a}}{\ell}\frac{2x}{y}<1 
		\Rightarrow  y>\frac{2z_{a}}{\ell} \quad(0\le x \le 1).
\label{constraint}
\end{equation}
This is in contrast with the boundary $y\sim 0$ in the integral in Eq. \eqref{start}. However, since $z_{a}\ll \ell$, the range $0<y<2z_{a}/\ell$ gives a subleading contribution to the integral and can safely be neglected.
For the reflection coefficient we hence obtain 
\begin{align}
	r^{\rm TE}\left(\omega,\frac{y}{2z_{a}}\right)
		&\approx \frac{1}{2}\int_{0}^{1}{\rm d}x~
				 \frac{\frac{2}{\pi}}{\sqrt{1-x^2}}
			     \frac{\imath\omega\frac{3\pi}{4}
			     	   \frac{\lambdabar_{\rm p}}{v_{\rm F}}
					   \frac{x^{3}}{y^{3}}
					   \left(\frac{2z_{a}}{\lambdabar_{\rm p}}\right)^{3}
					  }
		   	          {1-\imath\omega\frac{3\pi}{4}
		   	           \frac{\lambdabar_{\rm p}}{v_{\rm F}}
				 \frac{x^{3}}{y^{3}}
				 \left(\frac{2z_{a}}{\lambdabar_{\rm p}}\right)^{3}}\nonumber\\
		&\approx \imath\frac{3\pi}{4}\omega
				 \frac{ z_{a}}{v_{\rm F}}
			     \left(\frac{2z_{a}}{\lambdabar_{\rm p}}\right)^{2}
			     \int_{0}^{1}{\rm d}x~\frac{\frac{2}{\pi}}{\sqrt{1-x^2}}
				 \frac{x^{3}}{y^{3}}\nonumber\\
	    &=       \imath\frac{\omega}{y^{3}}\frac{ z_{a}}{v_{\rm F}}
				 \left(\frac{2z_{a}}{\lambdabar_{\rm p}}\right)^{2}.
\end{align}

Importantly, in the second line of the last expression we have approximated the denominator in a way which amounts to neglecting the dielectric function appearing in the denominator of Eq. \eqref{refTEApp}. This is possible since, from Eq. \eqref{epsapprox}, the dielectric function stays finite as long as $y>2z_{a}/\ell$ and the dominant values are for $y\sim 1$. The condition on the distance needs, however, a stronger constraint and $z_{a}\ll \ell$ must be replaced with 
\begin{equation}
	z_{a}\ll \ell \left[\frac{2}{3\pi}
			      \frac{\lambdabar_{\rm p}^{2}}{\ell^{2}}\right]^{1/3}.
\end{equation}
Inserting the previous result in Eq. \eqref{start} we obtain
\begin{align}
	\Delta^T\left(z_a,\omega\right)
		&\approx -4\imath\omega
				 \frac{\Phi^{T}(\omega)}{v_{\rm F}\lambdabar_{\rm p}^{2}}
 			     \int_{\frac{2z_{a}}{\ell}}^{\infty}{\rm d}y~
 			     \frac{e^{-y}}{y}\nonumber\\
		&\approx 4\imath\omega
				 \frac{\Phi^{T}(\omega)}{v_{\rm F}\lambdabar_{\rm p}^{2}}
				 \ln\left[\frac{2z_a}{\gamma'_{\rm E}\ell}\right],	 	   
 \label{final}
\end{align}
where $\gamma'_{\rm E}=e^{-\gamma_{\rm E}}$ with $\gamma_{\rm E}$ the Euler-Mascheroni constant.
Using $\varrho^{T}(z_a,\omega)=-\partial_{\omega}\mathrm{Im}[\Delta^{T}(z_{a},\omega)]$, the previous equation directly leads to Eq. \eqref{nonlocalmodedensity}.

The last steps in the above derivation are no longer valid if $\Gamma\to 0$. This limit is interesting for the evaluation of the entropy at small temperatures and therefore we focus only on the limit $|\omega|\to 0$. In this case $\ell \to \infty$, leading to the divergence of Eq. \eqref{final} or,  equivalently, in the density of modes in Eq. \eqref{nonlocalmodedensity}.  However, since the constraint in Eq. \eqref{constraint} includes now $y=0$, it is not possible to neglect the dielectric function in the denominator of Eq. \eqref{refTEApp}. 
A simplification is possible if we impose that
\begin{equation}
	y\gg \left[\frac{3\pi}{4}
			   \frac{|\omega|\lambdabar_{\rm p}}{v_{\rm F}}
		 \right]^{1/3}
		 \frac{2z_{a}}{\lambdabar_{\rm p}}
	  =  2\left[\frac{\hbar|\omega|}{E^{\rm nl}}\right]^{1/3},
\end{equation}
where $E^{\rm nl}$ is the characteristic energy of the eddy currents for the nonlocal description of the metal given in Eq. \eqref{lambdae}.
For $|\omega|\to 0$ the previous limit amounts to neglecting a subleading contribution in the integral in Eq. \eqref{start}.
We then obtain
\begin{align}
	\Delta^T\left(z_a,\omega\right)
		&\approx \frac{4}{3}\imath\omega
				 \frac{\Phi^{T}(0)}{v_{\rm F}\lambdabar_{\rm p}^{2}}
			     \ln\left[\left(\frac{2}{\gamma'_{\rm E}}\right)^{3}
			     		  \frac{\hbar|\omega|}{E^{\rm nl}}
			     	\right].
\label{final2}
\end{align}


\section{Entropy behavior at low temperature}
\label{Entropyanalysis}
In this Appendix, we analyze the Casimir entropy at low temperature. Our approach is based on an improvement of a method originally presented in \cite{Intravaia08}. 
As explained in Sec. \ref{EntropyPhenomenology}, the entropy is related to the behavior of the function $\Delta^{T}(z_{a}, \zeta)$, where $\zeta$ is a complex frequency.  
Dropping, for simplicity, the distance dependence from Eq. \eqref{entropy} we have
\begin{align}
\label{A1}
	\frac{\mathcal{S}(z_a,T)}{\pi k_{\rm B}}=
		&-\frac{d}{d\nu}\sideset{}{'}\sum_{n=0}^{\infty}
			\nu\Delta^{T}(\imath n \nu)\nonumber\\
		=&-\sum_{n=1}^{\infty}\left[\Delta^{T}(\imath n \nu)
			+\imath n \nu\Delta_{\zeta}^{T}(\imath n \nu)\right]\nonumber\\
		&-\frac{\Delta^{T}(0)}{2}
		 -\sum_{n=0}^{\infty}{'} 
			\nu\frac{\hbar \partial_{T}\Delta^{T}(\imath n \nu)}{2\pi k_{\rm B}}.
\end{align}
Here, the prime indicates that the first term of the sum has to be taken with a prefactor $1/2$ and we define $\Delta_{\zeta}^{T}(\omega)=[\partial_{\zeta}\Delta^{T}(\zeta)]_{\vert \zeta=\omega}$. Upon writing $\Delta^{T}(\imath n \nu)=n\Delta^{T}(\imath n \nu)-(n-1)\Delta^{T}(\imath n \nu)$ and after straightforward but rather cumbersome algebra, we can write the second line in the above expression as
\begin{align}
	&-\sum_{n=1}\imath n\nu
		\left[ \Delta_{\zeta}^{T}(\imath n \nu)
			+\frac{\Delta^{T}(\imath n \nu)
			-\Delta^{T}(\imath (n+1)\nu)}{\imath \nu}
		\right]\nonumber\\
	&=\sum_{n=1,m=2}n\Delta_{\zeta^{m}}^{T}(\imath n \nu)
		\frac{(\imath \nu)^{m}}{m!}\nonumber\\
	&=\oint_{C}\frac{dz}{2\pi\imath}\sum_{n=1,m=2}
		\frac{n(\imath \nu)^{m}}{(z-\imath n \nu)^{m+1}}\Delta^{T}(z),
\end{align}
where $C$ denotes a path that encloses counterclockwise the upper part of the complex $z$-plane. In the last step of the previous calculation we use that $\Delta^{T}(z)$ is analytic in the upper part of the complex plane. In the limit of $\nu\to 0$, we obtain
\begin{align}
	\sum_{n=1}\frac{n(\imath \nu)^{m}}{(z-\imath n \nu)^{m+1}}
		&=\sum_{n=0}
			\frac{(\imath\nu)^{m-1}(\imath[n+1]\nu)}{(z-\imath [n+1]\nu)^{m+1}}	
			\nonumber\\
		&\approx\int_{0}^{\infty}{\rm d}x
			\frac{(\imath \nu)^{m-2}[x+\imath \nu]}{(z- [x+\imath\nu])^{m+1}}
			\nonumber\\
		&=\frac{(\imath \nu)^{m-2}(z-\imath m \nu )}{m(m-1)(z-\imath\nu )^{j}}.
\end{align}
Then, the first two terms of the series in $m$ are
\begin{align}
	&\oint_{C}\frac{dz}{2\pi\imath}
		\left[\frac{(z-\imath 2 \nu )}{2(z-\imath\nu )^{2}}
			+\frac{\imath \nu(z-\imath 3 \nu )}{6(z-\imath\nu )^{3}}
		\right]
		\Delta^{T}(z)\nonumber\\
	&\approx \frac{1}{2} \Delta^{T} (\imath\nu )
		-\frac{1}{3}\imath\nu \Delta_{\zeta}^{T}(\imath \nu )
		\nonumber\\
	&= \frac{1}{2} \Delta^{T} (\imath\nu )
		-\frac{1}{3}\nu  \partial_{\nu}\Delta^{T}(\imath \nu ).
\end{align}
Collecting all the terms and approximating the rightmost sum in Eq. \eqref{A1} by an integral we obtain
\begin{align}
\label{Asymptotic}
	\frac{\mathcal{S}(T\to 0)}{\pi k_{\rm B}}
		\approx
		& \frac{\Delta^{T}(\imath\nu )-\Delta^{T}(0)}{2} 
		  -\frac{1}{3}\nu  \partial_{\nu}\Delta^{T}(\imath\nu )\nonumber\\
		& -\int_{0}^{\infty}{\rm d}\xi
		  \frac{\hbar \partial_{T}\Delta^{T}(\imath \xi)}{2\pi k_{\rm B}}.
\end{align}
Upon assuming that the function $\Delta^{T}(\imath\nu )$ is well-behaved in the limit $\nu\to 0$ (see the main text), Eq. \eqref{Asymptotic} can be further simplified as
\begin{align}
\label{Asymptotic2}
	\frac{\mathcal{S}(T\to 0)}{\pi k_{\rm B}}
		&\approx \frac{1}{6}\nu\varrho^{T}(0)
				 -\int_{0}^{\infty}{\rm d}\xi
				 \frac{\hbar \partial_{T}\Delta^{T}(\imath \xi)}{2\pi k_{\rm B}},
\end{align}
where we use that $\Delta^{T}(\omega)=\mathrm{Re}[\Delta^{T}(\omega)]+\imath \mathrm{Im}[\Delta^{T}(\omega)]$ (with the real and imaginary part being, respectively, even and odd in $\omega$) and that $\varrho^{T}(\omega)=-\partial_{\omega}\mathrm{Im}[\Delta^{T}(\omega)]$.
In most of the cases we can safely neglect the third term in Eqs. \eqref{Asymptotic} and \eqref{Asymptotic2}
\begin{align}
	\int_0^{\infty}{\rm d}\xi\partial_T\Delta^T(z_a,\imath\xi)
		=\frac{\partial\Gamma(T)}{\partial T}\partial_{\Gamma}
		 \int_0^{\infty}{\rm d}\xi\Delta^T(z_a,\imath\xi)
\end{align}
since it vanishes faster than linearly for vanishingly small temperatures.

Interestingly, the very same result of Eq. \eqref{Asymptotic2} can be obtained by regarding, in Eq. \eqref{FreeEnergy}, the expression $\coth\left[\hbar\omega/(2\kb T)\right]$  as a distribution.  Then, we can employ its asymptotic moment expansion \cite{Estrada90,Estrada02}
\begin{align}
	\coth\left[\tau\omega\right]
		\sim \mathrm{sgn}[\omega]
			 +\sum_{n=0}\frac{\mu_{n}(-1)^{n}\delta^{(n)}(\omega)}{n!\tau^{n+1}}		    \text{ for } \tau\to \infty,
\label{momemtExp}
\end{align}
where $\mu_{n}$ denotes the moments of the (regularized) distribution defined as
\begin{equation}
	\mu_{n}=
		\begin{cases}
			0											& n\text{ even}\\
			\int_{-\infty}^{\infty}{\rm d}\omega\, 
			\omega^{n}
			\left(\coth[\omega]-\mathrm{sgn}[\omega]
			\right)										& n\text{ odd}
		\end{cases}.
\end{equation}
In our case, $\tau=\hbar/(2 k_{\rm B}T)$ is the thermal coherence time, which indeed tends to infinity for $T\to 0$.
In order to recover Eq. \eqref{Asymptotic2}, we consider the first two terms of the sum in Eq. \eqref{momemtExp}, with $\mu_{1}=\pi^{2}/6$ and $\mathrm{Im}[\Delta^{T}(z_{a},0)]=0$. The previous expression also allows us to easily consider higher order terms and write
\begin{align}
\label{EntropyGeneral}
	&\frac{\mathcal{S}(z_a,T)}{k_{\rm B}}=\nonumber\\
	& \quad-\sum_{k=0}\frac{\mu_{2k+1}(k+1)\mathrm{Im}
		\Delta_{\omega^{2k+1}}(z_{a},0)}{(2k+1)!}
		\left[\frac{2 k_{\rm B}T}{\hbar}\right]^{2k+1}.
\end{align}


\section{Density of eddy currents}
\label{appendix1}
In this Appendix, we analyze the contribution to the density of mode $\eta^{T}(z_{a},\omega)$ stemming from the eddy currents. 
Starting from the definition in Eq. \eqref{modeeddy} we focus on the limit $\omega \to 0$ and consider first the case with a constant damping rate $\Gamma$. In order to simplify the calculation we consider the limit where the contribution of the eddy currents is more relevant, i.e. when $\Gamma$ is the smallest (but finite) frequency scale in the system. In this case we have that 
\begin{align}
	 \eta^{T}(z_a,0)
	 	\approx& \Phi^{T}(0)\mathcal{I}(2,2)
	 			 -\frac{\mu_0}{(8\pi)^{2}}\beta_{zz}^T(0)
	 			 \frac{\mathcal{I}(0,0)}{c^2}.
\end{align}
Here, we define the parameter-dependent integral
\begin{align}\label{I}
	\mathcal{I}
		&(i,j)=\nonumber\\
		&-\frac{8}{\pi}\text{Im}
			\int_{0}^{\Gamma}\frac{{\rm d}\xi}{\xi^{i}}
			\int_{\xi/c}^{\infty}{\rm d}\kappa~\kappa^{j} e^{-2\kappa z_a}
			\frac{\kappa-\sqrt{\kappa^2-\kappa^{2}_{\xi}}}
				 {\kappa+\sqrt{\kappa^2-\kappa^{2}_{\xi}}}.
\end{align}
The corresponding integrand becomes complex valued as soon as  $\kappa^2<\kappa^{2}_{\xi}=\xi/[D-\xi\lambdabar_{\rm p}^2]$, thus defining an upper bound for the $\kappa$-integration. Substituting $dy=d\kappa/\kappa_{\xi}$ and $d\kappa_{\xi}=D\kappa_{\xi}^3 d\xi/(2\xi^2)$, the integral reads 
\begin{align}
	\mathcal{I}(i,j)=
		&\frac{32}{\pi D}
			\int_{0}^{\infty}{\rm d}\kappa_{\xi}~\kappa_{\xi}^{j-2}
			\left(\frac{D\kappa^{2}_{\xi}}
					   {\kappa^{2}_{\xi}\lambdabar_{\rm p}^{2}+1}
			\right)^{2-i}\nonumber\\
		&\times\int_{\varphi(\xi)}^{1}{\rm d}y~y^{j+1}\sqrt{1-y^2}
				e^{-2y\kappa_{\xi} z_a},
\end{align}
where $\varphi(\xi)=\sqrt{(\xi\Gamma-\xi^2)/\omega_{\rm p}}\in[0,\Gamma/(2\omega_{\rm p})]$ is a symmetric positive function centered around $\Gamma/2$ and vanishing at $\xi=0,\Gamma$. In the limit of good conductors, $\Gamma/\omega_{\rm p}\ll 1$, we can approximate $\varphi(\xi)\approx 0$. The remaining integral can be evaluated exactly obtaining $\mathcal{I}(2,2)=[D z_{a}]^{-1}$, while $\mathcal{I}(0,0)/c^2\propto D/c^2$. Upon neglecting this subleading contribution, $\eta^{T}(z_a,0)$ becomes equivalent to the result reported in Eq. \eqref{lowfrequencies}.
Using the results of Appendix \ref{Entropyanalysis}, one can show that in this case the entropy contribution stemming from the eddy modes vanishes linearly with the temperature as $\mathcal{S}_{\rm e}=(\pi^2/3)(k_{\rm B}^2/\hbar)\eta_{\rm e}^0(z_a,0)T$. 

Consider now the case where the dissipation rate obeys a power-law dependence with regards to temperature, i.e. $\Gamma\propto T^m$ where $m\geq2$. In general, starting from Eq. \eqref{modeeddy}, after a partial integration, the entropy due to eddy modes can be written as
\begin{align}
	\mathcal{S}_{\rm e}=\frac{\hbar}{2\pi}\partial_T\int_0^{\infty}
		&{\rm d}\omega~\coth\left[\frac{\hbar\omega}{2\kb T}\right]\nonumber\\
		\times\int_{0}^{\Gamma}&{\rm d}\xi ~
			\frac{\omega}{\xi^2+\omega^2}\text{Im}\Delta^T(z_a,-\imath\xi+0^+).
\end{align}
We make the change of variables $\xi =x\Gamma$, $x<1$, $\omega=z \Gamma$ and obtain
\begin{align}
	\mathcal{S}_{\rm e}=\frac{\hbar}{2\pi}\partial_T&\left\{ \Gamma \int_{0}^{1}{\rm d}x\, ~ \text{Im}\Delta^T(z_a,-\imath x \Gamma+0^+)\right.
		\nonumber\\
		\times&\left.\int_0^{\infty}{\rm d}z~\coth\left[z\frac{\hbar\Gamma}{2\kb T}\right]\frac{z}{x^2+z^2}
			\right\}.
\end{align}
Since only the thermal contribution is relevant and the integral is convengent, $z\lesssim 1$ and for $\Gamma/T \to 0$ the cotangent hyperbolic can be approximated with the inverse of its argument. The resulting integral in $z$ can be performed and as a result, in the case of a perfect crystal, we obtain for the low-temperature limit of the eddy currents' contribution to the entropy 
\begin{align}
	\mathcal{S}_{\rm e}
		=\frac{\kb}{2}\int_{0}^{1}{\rm d}x~   
		\left.\frac{\text{Im}
			\Delta^0(z_a,-\imath\Gamma x+0^+)}{x}
		\right|_{\Gamma\rightarrow 0},
\end{align}
which is independent of the dissipation rate $\Gamma$ [see Eq. \eqref{I}]. Using Eq. \eqref{start} and performing the $y$-integration, we eventually obtain
\begin{align}\label{entropyeddy}
	\mathcal{S}_{\rm e}
	=4\pi\kb
		\frac{\Phi^{0}(0)}{z_a^3}F
			 \left(\frac{z_a}{\lambdabar_{\rm p}}\right),
\end{align}
where $F(x)$ is the function defined below Eq. \eqref{entropydefect}. Equation \eqref{entropyeddy} confirms the argument outlined below Eq. \eqref{FreeEnergyEddy}, i.e. $\mathcal{S}_{\rm e}\rightarrow\mathcal{S}_0$ ($T\rightarrow 0$). 


\section{The nonlocal branch-point}
\label{Branchpoint}
In this Appendix, we study the characteristics of the nonlocal branch-point $\xi_{0}^{\rm nl}(p)$ defined as the purely imaginary solution of $\epsilon_t(\omega,p)\omega^2/c^2-p^2=0$.
To this end it is convenient to change the variables according to $\omega=-\imath\xi=-\imath\Gamma(1-s)$, so that real values of $\xi$ correspond to real values of $s$. The passivity of our system implies that $s<1$ and from comparison with the local case we expect that $s>0$. After some manipulations, using the expressions in Eqs. \eqref{permittivities}, the zeros of the argument of the square root in the denomination of Eq. \eqref{SI2} can be found by solving
\begin{equation}
	s^{3}\frac{\lambdabar_{\rm p}^{2}}{\ell^{2}}
	+\frac{\Gamma^{2}}{\omega_{\rm p}^{2}}s(1-s)^{2}x^2
	-(1-s)g_t\left(x\right)x^2
	=0.
\label{eqS}
\end{equation}
For finding $\xi_{0}^{\rm nl}(p)$ we are interested in the real solution of the previous equation only,
where $x=s/(p\ell)>0$ and the function $g_t(x)$ is given by
\begin{align}
	g_t(x)
	=\frac{3}{2}\left[(1+x^{2})x \, \arccot\left[x\right]-x^{2}\right]
	\approx 
	\begin{cases}
		\frac{3\pi}{4}\abs{x}		 & x\ll1\\
		1 							 & x\gg 1
		\end{cases}.
\end{align}
If we look at $s$ and $x$ as two independent variables, Eq. \eqref{eqS} can be solved exactly for $s(x)$. But for our purpose it is sufficient to consider the reduced cubic equation obtained by neglecting the term  proportional to $\Gamma^{2}/\omega_{\rm p}^{2}$ which usually is quite small in metals:
\begin{equation}
	s^{3}
	+s\frac{\ell^{2}}{\lambdabar_{\rm p}^{2}}g_t\left(x\right)x^2
	-\frac{\ell^{2}}{\lambdabar_{\rm p}^{2}}g_t\left(x\right)x^2
	=0. 
\end{equation}
Since $g_t\left(x\right)x^2\ge0$, the real solution of the previous equation is given by
\begin{equation}
	s_{0}(x)
	=2x\frac{\ell}{\lambdabar_{\rm p}}
	 \sqrt{\frac{g_t\left(x\right)}{3}}
	 \sinh\left[\frac{1}{3}\arcsinh
	 			\left(\frac{3 }{2x}
	 				  \frac{\lambdabar_{\rm p}}{\ell}\sqrt{\frac{3}{g_t
	 				  \left(x\right)}}
	 			\right)
	 		\right].
\label{simpleEq}
\end{equation}
We can then describe the behavior of $\xi^{\rm nl}_{0}(p)$ in terms of the parametric solution
\begin{gather}\label{parametricXi}
	\xi^{\rm nl}_{0}(p)\approx
		\begin{cases}
			\Gamma[1-s_{0}(x)]\\
			p=\frac{s_{0}(x)}{x\ell}
		\end{cases}.
\end{gather}
Specifically, we find that $s_{0}(0)=0\le s_{0}(x)\le 1=s_{0}(\infty)$ indicating that for the branch-point we have $0\le \xi_{0}^{\rm nl}(p)\le \Gamma$, where the value $\xi_{0}^{\rm nl}(p)=0$ corresponds to $x\to \infty$ and $p=0$. The relation $\xi_{0}^{\rm nl}(p)=\Gamma$ is fulfilled for 
\begin{equation}
	p=\frac{1}{\ell}
	\left[\frac{3\pi}{4}
		  \frac{\ell^{2}}{\lambdabar_{\rm p}^{2}}
	\right]^{\frac{1}{3}}.
\end{equation}
This also is the maximum value that the variable $p$ can take. If $\ell \gg \lambdabar_{\rm p}$ then $x\ll 1$ ($v_{\rm F}/c\gg \Gamma/\omega_{\rm p}$) and the solution takes the form given in Eq. \eqref{xinl}. This can be seen directly from Eq. \eqref{simpleEq} which now reads
\begin{align}
	&1+s\frac{\ell^{2}}{\lambdabar_{\rm p}^{2}}
		\frac{3\pi}{4}\left(\frac{1}{p\ell}\right)^{3}
		-\frac{\ell^{2}}{\lambdabar_{\rm p}^{2}}
		\frac{3\pi}{4}\left(\frac{1}{p\ell}\right)^{3}\approx0\nonumber\\
	&\Rightarrow s \approx 1- \frac{4}{3\pi}p^{3}\ell\lambdabar_{\rm p}^{2}.
\end{align}
As a result, the corresponding frequency is 
\begin{align}
	\xi^{\rm nl}_{0}(p)\approx\frac{4}{3\pi}p^{3}v_{\rm F}\lambdabar_{\rm p}^{2}
\end{align}
which coincides with the expression given in Eq. \eqref{xinl}.



\end{document}